\documentclass{aa}  

\usepackage{natbib}
\bibpunct{(}{)}{;}{a}{}{,} 
\usepackage{graphics}   
\usepackage{graphicx}   
\usepackage{epstopdf}
\usepackage[varg]{txfonts}
\usepackage{pdflscape}
\usepackage{hyperref}   
\usepackage[flushleft]{threeparttable}
\usepackage{longtable}
\usepackage{caption}
\usepackage{balance}
\usepackage{color}
\usepackage{pdfpages}
\usepackage{amsfonts,amssymb,latexsym}
\usepackage[normalem]{ulem}

\hypersetup{
    bookmarks=true,         
    colorlinks=true,       
    linkcolor=blue,          
    citecolor=blue,        
    filecolor=blue,      
    urlcolor=blue           
}

\def\ga{\,\,\raise0.14em\hbox{$>$}\kern-0.76em\lower0.28em\hbox{$\sim$}\,\,}
\def\la{\,\,\raise0.14em\hbox{$<$}\kern-0.76em\lower0.28em\hbox{$\sim$}\,\,}

\def\iso#1{$^{#1}$}
\def\Msun{$M_{\odot}$}
\def\Lsun{$L_{\odot}$}
\def\msun{M_{\odot}}
\def\pg{$p,\gamma$}
\def\ng{$n,\gamma$}
\def\cm3{cm$^{-3}$}
\def\an{$\alpha,n$}
\def\np{$n,p$}
\def\na{$n,\alpha$}
\def\pa{$p,\alpha$}

\def\chem#1#2{$\mathrm{^{#2}\kern-0.8pt#1}$}
\def\reac#1#2#3#4#5#6{$\mathrm{\, ^{#2}\kern-0.8pt{#1}\, ({#3}\, ,{#4})\, {}^{#6}\kern-0.8pt{#5}\, }$}
 
\def\conv{\mathrm{conv}} 

\def\nuc{\mathrm{nuc}} 
\def\min{\mathrm{min}} 
\def\max{\mathrm{max}} 
\def\be{\begin{equation}} 
\def\ee{\end{equation}}
\def\beqy{\begin{eqnarray}}
\def\eeqy{\end{eqnarray}}
\def\bmlet{\begin{mathletters}}
\def\emlet{\end{mathletters}}
\def\x{{\cal{X}}}

\begin{document}

\title{The intermediate neutron capture process}
\subtitle{I. Development of the i-process in low-metallicity low-mass AGB stars}

\author{A. Choplin   
\and L. Siess
\and S. Goriely}
\offprints{arthur.choplin@ulb.ac.be}

\institute{
Institut d'Astronomie et d'Astrophysique, Universit\'e Libre de Bruxelles,  CP 226, B-1050 Brussels, Belgium
}

\date{Received --; accepted --}

\abstract
{
Results from observations report a growing number of metal-poor stars showing an abundance pattern midway between the s- and r-processes.
These so-called r/s-stars raise the need for an intermediate neutron capture process (i-process), which is thought to result from the ingestion of protons in a convective helium-burning region, but whose astrophysical site is still largely debated.
}
{
We investigate whether an i-process during the asymptotic giant branch (AGB) phase of low-metallicity low-mass stars can develop and whether it can explain the abundances of observed r/s-stars. 
}
{
We computed a 1 \Msun\ model at [Fe/H]~$=-2.5$ 
with the stellar evolution code {\sf STAREVOL}, using a nuclear network of 1091 species (at maximum) coupled to the transport processes. 
The impact of the temporal and spatial resolutions on the resulting abundances was assessed. 
We also identified key elements and isotopic ratios that are specific to i-process nucleosynthesis and carried out a detailed comparison between our model and a sample of r/s-stars.
}
{
At the beginning of the AGB phase, during the third thermal pulse, the helium driven convection zone is able to penetrate the hydrogen-rich layers. The subsequent proton ingestion leads to a strong neutron burst with neutron densities  of $\sim 4.3 \times 10^{14}$~\cm3 at the origin of the synthesis of i-process elements. The nuclear energy released by proton burning in the helium-burning convective shell strongly affects the internal structure: the thermal pulse splits and after approximately ten years the upper part of the convection zone merges with the convective envelope. The surface carbon abundance is enhanced by more than 3 dex. This leads to an increase in the opacity, which triggers a strong mass loss and prevents any further thermal pulse.
Our numerical tests indicate that the i-process elemental distribution is not strongly affected by the temporal and spatial resolution  used to compute the stellar models, but typical uncertainties of $\pm 0.3$~dex on individual abundances are found.
We show that specific isotopic ratios of Ba, Nd, Sm, and Eu can represent good tracers of i-process nucleosynthesis. Finally, an extended comparison with 14 selected r/s-stars show that the observed composition patterns can be well reproduced by our i-process AGB model.
}
{
A rich i-process nucleosynthesis can take place during the early AGB phase of low-metallicity low-mass stars and explain the elemental distribution of most of the r/s-stars.
}

\keywords{nuclear reactions, nucleosynthesis, abundances -- stars: AGB and post-AGB}

\titlerunning{Development of the i-process in low-metallicity low-mass AGB stars}

\authorrunning{A. Choplin et al. }

\maketitle


\section{Introduction}
\label{sect:intro}
Elements heavier than iron are mainly produced by neutron capture reactions \citep[e.g., the review of ][]{arnould20}. 
Two main neutron capture processes are known to be responsible for about half of the trans-iron elements in the Universe: the slow and rapid neutron-capture process (s- and r-process, respectively).

The r-process is characterized by neutron densities $N_n$ in excess of $10^{20}$\,\cm3, operating during a  timescale on the order of a second. 
The site for r-process nucleosynthesis remains uncertain, but a strong candidate is the merging of two neutron stars  \citep[e.g.,][]{arnould07,goriely11a,goriely11b,wanajo14,just15}. This scenario is now supported by the observation of a neutron star merger event \citep{abbott17} whose electromagnetic counterpart is compatible with radioactive powering from r-process nuclides \citep{smartt17, pian17}.
Other astrophysical sites for the r-process include collapsars \citep[e.g.,][]{siegel19} and magneto-rotational supernovae \citep[MRSNe; e.g.,][]{winteler12, nishimura15}, although unrealistically large pre-collapse magnetic fields may be required to power an efficient r-process nucleosynthesis in MRSNe \citep{mosta18}.

The s-process is associated with low neutron densities ($N_n \simeq 10^6-10^{10}$\cm3) and is usually separated into a weak and main component.
The weak s-process, mostly powered by the reaction \iso{22}Ne(\an), is responsible for the production of elements with atomic number $A \lesssim 90$. It occurs in the core helium-burning of massive stars \citep[e.g.,][]{langer89, prantzos90} and to a smaller extent in their carbon-burning shell \citep[e.g.,][]{raiteri91b,the07}. 
Rotation was shown to significantly boost the weak s-process \citep[e.g.,][]{pignatari08,frischknecht12}, but its quantitative effect on the nucleosynthesis is still debated \citep[][]{frischknecht16, choplin18, limongi18, banerjee19}.
The main s-process, which is able to produce elements with $A \gtrsim 90$, was recognized to occur in low- and intermediate-mass stars during the asymptotic giant branch (AGB) phase \citep[e.g.,][]{gallino98, herwig05, cristallo11, karakas14}. In this case the \iso{13}C(\an) reaction is the main neutron source although \iso{22}Ne(\an) can also be activated, especially in low-metallicity and massive AGB stars \citep{lugaro12}.
Important uncertainties remain regarding the formation mechanism of the \iso{13}C pocket, which mainly determines the resulting nucleosynthesis \citep[e.g.,][]{battino19,vescovi20}, the dredge-up efficiency or the mass loss \citep[Sect. 4 in][for a discussion about these uncertainties, and reference therein]{karakas14}.

Other nuclear processes also contribute to forging heavy elements. 
The p-process \citep[e.g., the review of][]{arnould03} and $\nu p$-process \citep{frohlich06} can reproduce the solar abundances of the neutron deficient isotopes (or p-nuclides) such as \iso{92}Mo or \iso{94}Mo. Such processes can occurs in the oxygen-burning shell of massive stars shortly before collapse, during the supernova (SN) explosion \citep{rayet95,rauscher02,frohlich06,wanajo11}, and during type Ia SNe \citep{travaglio15}. 
A weak r-process can occur in the neutrino-driven wind of core-collapse supernovae. This would build elements such as Sr, Y, or Zr \citep{arcones13}.
A short neutron burst in the helium shell of massive stars when the supernova shock wave passes through was also recognized and referred to as the n-process \citep[e.g.,][]{blake76, thielemann79, meyer04, choplin20}.

In addition, an intermediate neutron capture process (i-process) with neutron concentrations intermediate between the s- and r-processes ($N_n \approx 10^{12}-10^{16}$\cm3) was identified. 
\cite{cowan77} first pointed out the possibility of an i-process in red giant stars.
The hypothesis of an i-process later gained a lot of attention with the discovery of so-called r/s-stars, whose abundances are difficult to reproduce solely with an s-process, an r-process, or a combination of both \citep[e.g.,][]{jonsell06, lugaro12,dardelet14,roederer16}.
During an i-process event, the neutrons are released when protons are ingested in a convective region powered by helium burning. In this event, the protons are entrained by the convective flow to regions of higher temperature ($T > 10^8$K) where they are depleted via the reaction \iso{12}C(\pg)\iso{13}N. The subsequent beta decay of \iso{13}N to \iso{13}C is followed by the reaction \iso{13}C(\an)\iso{16}O, which produces high neutron densities up to $N_n \approx 10^{15}-10^{16}$\cm3 \citep[e.g.,][]{siess02,campbell08,herwig11}.
We list below the sites that have been suggested to host such a unique nucleosynthesis and that may therefore be at the origin of the surface enrichment in r/s-stars. 

One possible site is the early AGB phase of metal-poor low-mass stars, in which the entropy barrier between the hydrogen- and helium-rich zones can be surmounted by the energy released by a thermal pulse \citep{fujimoto00,chieffi2001,siess02,iwamoto04,cristallo09b,suda10,cristallo16}. The penetration of the helium driven convection zone in the hydrogen-rich layers is favored in metal-poor stars where composition gradients are reduced as a result of the absence of heavy elements. 
Another site is the core helium flash of very low-metallicity low-mass stars \citep{fujimoto90,fujimoto00,schlattl2001,suda10,campbell10, cruz13}. 
A third site is that of the very late thermal pulses of post-AGB stars \citep{herwig11}. 
Rapidly accreting carbon-oxygen (C-O) or oxygen-neon (O-Ne) white dwarfs (RAWDs) in close binary systems could also host i-process nucleosynthesis \citep{denissenkov17,denissenkov19}. In this scenario, the helium shell formed by the combustion of the accreted hydrogen material becomes thermally unstable. As a consquence of convection, protons from the accreted hydrogen-rich envelope may reach the helium-burning shell and lead to the production of an intense neutron burst.
A fifth site is that of the super-AGB stars (7~$\msun \lesssim M_{\rm ini} \lesssim$~10~$\msun$). \cite{jones16b} show that at low metallicity ($Z\leq 10^{-3}$), a 7~$\msun$ model can experience a proton ingestion episode during the thermally pulsing super-AGB phase leading to i-process nucleosynthesis. 
Another potential site in super-AGB stars is during the so-called dredge-out when, at the end of carbon burning, a helium-driven convective shell merges with the descending convective envelope \citep{siess2007}.
A sixth site is the helium shell of very low- or zero-metallicity massive stars ($M_{\rm ini} > $~10~$\msun$). These stars can experience a proton ingestion episode during central carbon burning (or later stages) and reach neutron densities of $10^{13} - 10^{15}$\cm3  \citep{banerjee18, clarkson18, clarkson20}.

Few studies have explored the i-process in stellar models under realistic conditions. Also, the dependence of this process on initial mass, metallicity, mixing (e.g., overshoot and rotation), or nuclear reactions remains to be assessed. 
Among past studies, we mention \cite{cristallo09b}, who investigated the early-AGB phase of a 1.5\Msun{} at [Fe/H]~$=-2.4$. with a network of $\sim 700$~isotopes.
Using a network of 320 species, \cite{campbell10} calculate the nucleosynthesis by post-processing in a 1 \Msun\ model at [Fe/H]~$=-6.5$, during the helium flash. 
\cite{denissenkov19} study the i-process nucleosynthesis by post-processing seven RAWDs models with $-2.6<$~[Fe/H]~$<0$ and using a network of $\sim 1000$ isotopes.
We also mention that existing study, aiming to compare predicted and observed surface abundances of r/s stars, did not use realistic stellar evolution calculations.
\cite{hampel16,hampel19} carry out detailed comparisons with r/s stars, but consider a one-zone model at constant temperature, density, and neutron density.

In this first paper of a series, 
we investigate how the i-process develops in a low-metallicity low-mass stellar model, its dependence on the numerics, the nuclear and chemical signatures of the i-process, and we compare our model predictions with abundances in r/s-stars. 
The specificity of this work is to consider for the first time AGB models with both high spatial resolution (up to $\sim 3000$ shells) and a large nuclear network (1091 species), coupled to the mixing processes at work during stellar evolution.
An analysis of the nucleosynthesis in stars of different masses and metallicities, as well as the effects of mixing processes such as overshooting or thermohaline, are left for future studies.

Sect.~\ref{sect_mod} describes the input physics. Sect.~\ref{sect_mix} investigates in detail the proton ingestion taking place in a fiducial 1~$\msun$ model with [Fe/H]~$=-2.5$. A sensitivity study of the nucleosynthesis is done in Sect.~\ref{sect_sens}. Differences between the s-, r-, and i-process abundance distributions are highlighted in Sect.~\ref{sect_distrib}. Finally, comparisons to observations are carried out in Sect.\ref{sect_comp} and conclusions are given in Sect.~\ref{sect_conc}.


\section{Input physics and numerical aspects}
\label{sect_mod}

The present calculations are based on the stellar evolution code  {\sf STAREVOL},  the description of which can be found in \cite{siess00}, \cite{siess06}, and \cite{goriely18c} and references therein. In present computations, the mixing length parameter $\alpha$ is set to 1.75 and we adopt the \cite{asplund09} solar mixture.
We take into account the change in  opacity due to the formation of molecules when the star becomes carbon rich, as prescribed by \cite{marigo02}. For the mass-loss rate, we use the \cite{reimers75} prescription with $\eta_R = 0.4$ from the main sequence up to the beginning of the AGB and then switch to the \cite{vassiliadis93} rate. We adopt gray atmosphere surface boundary conditions and integrate the stellar structure equations in one shot from the center up to an optical depth $\tau\approx 10^{-3}$. 
The following sections describe in more detail specific aspects of the modeling relevant to this study.

\subsection{Chemical transport}
\label{sect_over}

During proton ingestion, the mixing timescales becomes comparable to the nuclear timescale associated with the energetic reaction \iso{12}C(\pg). In these circumstances, nucleosynthesis cannot be computed independently of chemical transport. For this reason, we only consider the diffusive mixing scheme (DM) described in \cite{goriely18c}, where a diffusion equation is used to simulate the transport of chemicals, that is,  
\begin{equation}
\frac{\partial X_i}{\partial t}= \frac{\partial}{\partial m_r}  \left[\,(4 \pi r^2 \rho)^2\,D\ \frac{\partial X_i}{\partial m_r}\right] + \frac{\partial X_i}{\partial t}\bigg|_\nuc \ ,
\label{eq_dif}
\end{equation}
where $D = D_{\conv}+D_{\rm mix}$ is the sum of the diffusion coefficients associated with convection and other mixing mechanisms (such as overshoot), respectively. The last term in Eq.~(\ref{eq_dif}) accounts for the change in composition due to nuclear burning. 
In our study, from the onset of the AGB phase, the nucleosynthesis and diffusion equations are solved simultaneously once the structure has converged.
Also, in the present study, no additional mixing mechanism is included ($D_{\rm mix}=0$) in contrast to the s-process study performed in \cite{goriely18c}. The impact of such additional mixing mechanisms on the i-process nucleosynthesis will be studied in a forthcoming paper.

\subsection{Nuclear reaction network}
\label{sect_nucnet}
In this paper we consider two different networks of 411 and 1091 species (referred to as small and large networks, respectively), which are both coupled to a diffusion equation and solved once the structure has converged. 
The first network is described in \cite{goriely18c} and includes 411 nuclei between H and Po. The second network is extended to neutron-rich nuclei up to neptunium (Np) and comprises 1091 nuclei linked by with some  2104 nuclear ($n$-, $p$-, $\alpha$-captures, and $\alpha$-decays), weak (electron captures, $\beta$-decays) and electromagnetic interactions. 
The large reaction network involves all species with a decay half-life typically larger than 1~s and can consequently describe neutron capture processes characterized by irradiation of neutron densities up to a maximum of about $10^{17}$\cm3.

Nuclear reaction rates were taken from the Nuclear Astrophysics Library of the Free University of Brussels\footnote{available at http://www-astro.ulb.ac.be/Bruslib} \citep{arnould06} and include the latest experimental and theoretical cross sections through the interface tool NETGEN \citep{Xu13}. In particular, all the charged-particle-induced reaction rates of relevance in the hydrogen and
helium-burning calculations were taken from the NACRE and NACRE-II evaluations \citep{angulo99,xu13b} and the STARLIB library \citep{sallaska13}.
When not available experimentally, the cross sections were calculated within the statistical Hauser-Feshbach model with the TALYS reaction code \citep{goriely08a}. The TALYS calculations were also used systematically to deduce the stellar rates  from the laboratory neutron capture cross sections by
allowing for the possible thermalization of low-lying states in the target nuclei.  We note that at low temperature, the non-thermalization of the isomeric state of \iso{26}Al,\iso{85}Kr,\iso{115}In,\iso{176}Lu,  and \iso{180}Ta is introduced explicitly in the reaction network \citep{kappeler89,nemeth94}. The temperature- and density-dependent $\beta$-decay and electron capture rates in stellar conditions were taken from \citet{takahashi87} with the update of  \citet{goriely99a}.
The $(n,\alpha)$ reactions and $\alpha$-decays are also introduced when relevant, in particular between Bi and Np isotopes. 

We computed our models with the small network (411 nuclei) up to the proton ingestion event. Just before, we switched to the large network to follow the full i-process nucleosynthesis induced by the high neutron density ($N_n > 10^{14}$~\cm3, see Sect.~\ref{sect_mix}). Nevertheless, as discussed in Sect.~\ref{sect_netw}, the small network already gives a very good approximation of the i-process nucleosynthesis, hence this network can be used to obtain a first good guess of the full nucleosynthesis.


\section{Proton mixing in a 1~\Msun{}, [Fe/H]~$=-2.5$ stellar model}
\label{sect_mix}

We consider in this section a fiducial 1~\Msun~star computed with a metallicity [Fe/H]~$=-2.5$ ($Z=4.3\times10^{-5}$). 
This is our reference model (M$\epsilon 8 \alpha 10$ in Table~\ref{table:1}), which is taken as a reference when investigating the effects of temporal and spatial resolution in Sect.~\ref{sect_sens}. 
The model is computed with the small network (411 nuclei) except during the proton ingestion episode, at which point we switched to the large network (1091 nuclei).
We describe the structure and nucleosynthesis evolution of this model during the proton ingestion event below.

\begin{figure}[t]
\includegraphics[width=\columnwidth]{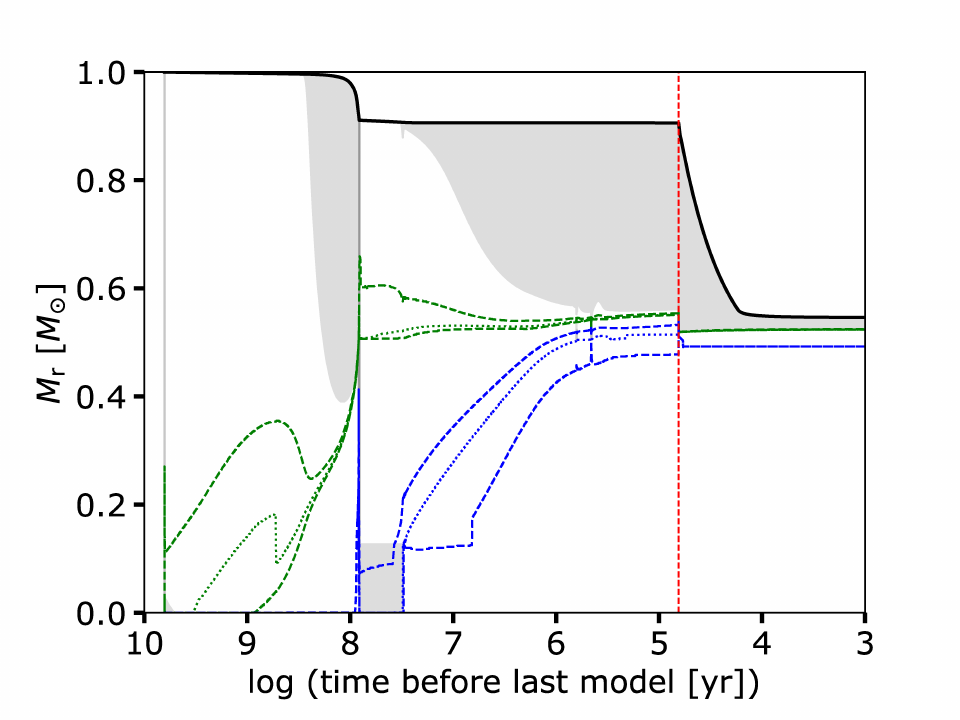}
\caption{
Kippenhahn diagram of the 1~\Msun\ stellar model at [Fe/H]~$=-2.5$ discussed in this paper. 
Time is counted backward from the last computed model. Convective regions are shaded gray. 
The vertical dashed red line indicates when the main proton ingestion event takes place (during the third thermal pulse). 
The dotted green and blue lines correspond to the maximum nuclear energy production by hydrogen and helium burning, respectively. The dashed green and blue lines delineate the hydrogen and helium-burning zones; more specifically they outline the zones in which the production of energy by hydrogen or helium burning exceeds 10~erg~g$^{-1}$~s$^{-1}$.
}
\label{fig_kip1}
\end{figure}

\begin{figure}[h!]
\includegraphics[width=\columnwidth]{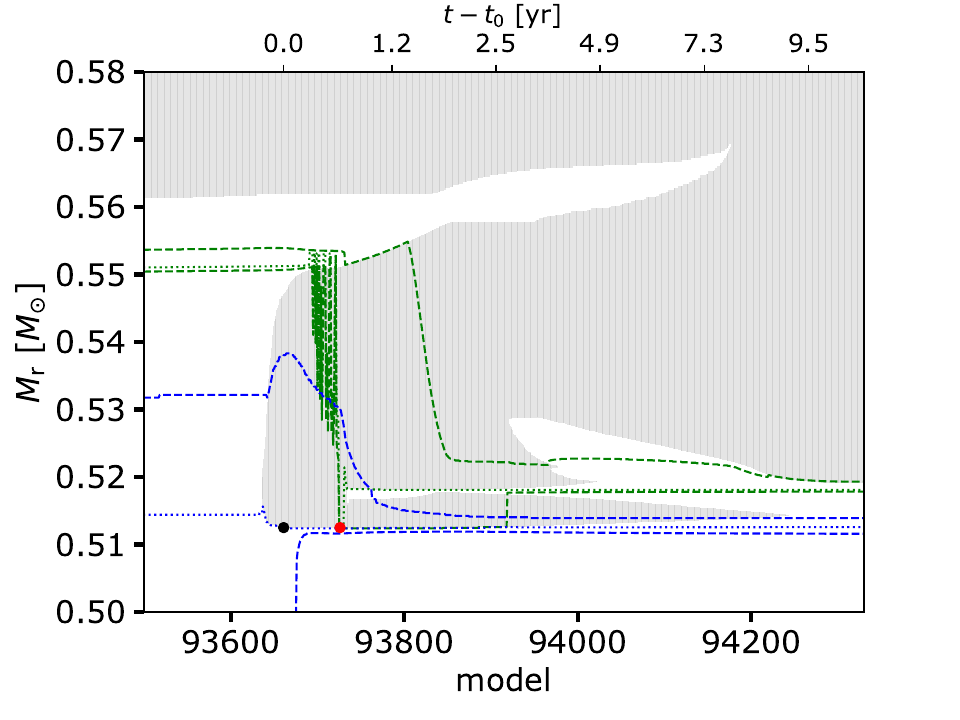}
\includegraphics[width=\columnwidth]{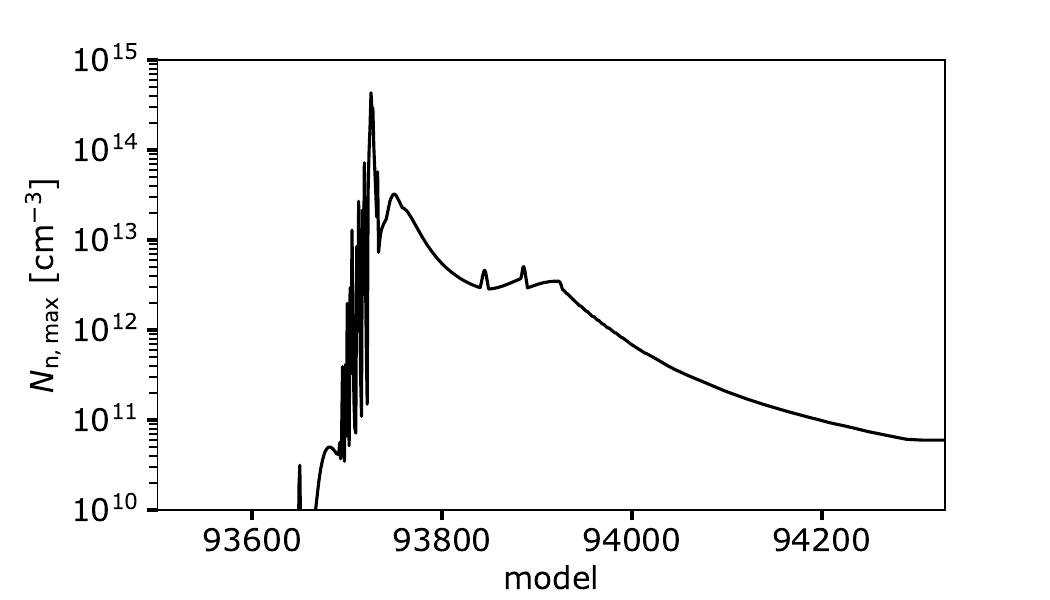}
\caption{Structure and maximum neutron density during the third instability. \textit{Top:} Kippenhahn diagram of the same model as in Fig.~\ref{fig_kip1} around the proton ingestion event. 
The red circle corresponds to the time and location where the neutron density is maximal. The chemical composition at the time and location of the two colored circles are shown in Fig.~\ref{fig_ab1}. 
The time is indicated at specific moments (it does not run linearly) on the top axis, counted from $t_0$, which is the time corresponding to the black circle. The three important steps in this event are the maximal neutron density (red circle, $t-t_0 = 1.182$~yr), the splitting of the convective helium shell ($t-t_0 = 1.184$~yr) and the merging of the helium shell with the envelope ($t-t_0 = 7.956$~yr). 
\textit{Bottom:} Evolution of maximum neutron density found in the stellar model at a given time (or equivalently for a given evolutionary model number).
}
\label{fig_kip2}
\end{figure}

\subsection{Structural evolution}\label{strucevol}

\subsubsection{pre-AGB evolution}

The evolution up to the beginning of the TP-AGB phase is very standard (Fig.~\ref{fig_kip1}). Core hydrogen burning proceeds radiatively and the star leaves the main sequence at an age of 5.84 Gyr. Degenerate core ignition occurs off-center at a mass coordinate $M_r = 0.22$~$\msun{}$ when the star is 6.28 Gyr old. At the end of core helium burning, when the convective envelope has reached its deepest extent during the second dredge-up (at an age of 6.36 Gyr), a C-O core mass of 0.50 \Msun{} is formed.

\subsubsection{AGB phase: First instabilities}

At the beginning of the TP-AGB phase, two weak convective instabilities develop in the helium-burning shell with an helium luminosity of $3 \times 10^4$ ($2 \times 10^6$) \Lsun\ for the first (second) instability.
During the second instability, the energy released is not sufficient to overcome the entropy barrier between the hydrogen and helium layers, but the top of the convective pulse nevertheless reaches the very bottom of the hydrogen shell and engulfs a small amount of hydrogen. This does not lead to a noticeable increase in the hydrogen luminosity and hardly alters the AGB star structure. This first weak proton ingestion episode however activates a weak i-process nucleosynthesis event (see Sect.~\ref{pulse_nuc}).

\subsubsection{AGB phase: The third instability}

During the third instability, the second and main proton ingestion event occurs (denoted by the vertical dashed red line in Fig.~\ref{fig_kip1}). 
The ingestion itself lasts for about one year and dramatically alters the subsequent AGB evolution. 
The top panel of Fig.~\ref{fig_kip2} shows a zoom around the third instability, where the main proton ingestion event occurs.
During this sequence, the evolutionary timestep was forced to not exceed $0.02$~yr. We noticed that allowing timesteps that are too large can lead to missing the ingestion event (see also Sect.~\ref{subsect_temporal} for additional discussion on the numerics).
Such proton ingestion phenomena were already identified in 1~\Msun\ models at [Fe/H]~$<-2$ by \cite{fujimoto00} and \cite{iwamoto04}. The penetration of the helium-burning shell in the hydrogen-rich layer is possible at low metallicity because the entropy barrier between the two zones is small and can be overcome by the energy released in the helium-burning shell.
Once the protons are ingested in the unstable helium-burning region, they are rapidly destroyed as they are advected downward by convective motions. 
The neutron density reaches a few $10^{14}$~\cm3\ (bottom panel of Fig.~\ref{fig_kip2}, cf. Sect.~\ref{pulse_nuc} for further details). 
This leads to a quick production of trans-iron elements that are mixed upward, up to $M_{\rm r} \sim 0.55 - 0.56$~\Msun. The nuclear energy production rate has reached its maximum 
where the timescale associated with the reaction \iso{12}C(p,$\gamma$) becomes similar to the convective timescale. At that location, energy accumulates, a temperature inversion develops, and the helium-burning convection zone finally splits (around model 93700 and $M_{\rm r} \sim 0.515$~\Msun\ in Fig.~\ref{fig_kip2}). At that time the luminosity generated by hydrogen burning reaches $10^{10}$ \Lsun{} and the timestep drops below $10^{-5}$~yr. In our simulations, it is important to stress that the maximum neutron density (red dot in Fig.~\ref{fig_kip2})  is reached before the convective helium-burning shell splits.

\subsubsection{AGB phase: Final stages}

About ten years after the proton ingestion event, the convective pulse merges with the envelope (around model 94200 in Fig.~\ref{fig_kip2}). It produces a strong surface enrichment in both light and heavy elements (cf. Sect.~\ref{pulse_nuc}). In particular, the surface \iso{12}C abundance rises by 3.1 dex. This has the effect of increasing the opacity of the outer layers, which in turn dramatically expands the outermost stellar layers and triggers intense mass loss ($\dot{M} \sim 10^{-5}$~$M_{\odot}$~yr$^{-1}$, after the red dashed line in Fig.~\ref{fig_kip1}).
The quick removal of the stellar envelope (in about $6 \times 10^{4}$~yr) prevents any further thermal pulse. We note that this may not be the case for more massive AGB stars because the enrichment is lower (larger dilution factors) and the opacity increase may not be as dramatic. Also, the shorter inter-pulse period may allow the development of helium shell instabilities before the total removal of a larger envelope. 
We note that the AGB models in \cite{cristallo09b, cristallo16} follow a normal AGB phase with thermal pulses after the ingestion episode. Their models are more massive (1.3 and 1.5 \Msun), which, as discussed above, may help the development of pulses after the ingestion. 
The surface CNO enrichment right after the ingestion is $\sim 0.5-1$~dex lower in their models \citep[at least in][their figure 6, top panel]{cristallo09b}, which yield lower opacities, radii, and mass-loss rates, and hence favor the appearance of pulses.
Finally, the difference in the input physics considered in \cite{cristallo09b, cristallo16} and in the present work (especially the different treatment of molecular opacities) could also play an important role in understanding the differences.\\

We note that the spiky location of the nuclear energy production by hydrogen-burning (green lines around model 93700) is due to the discrete mass progression of the top of the convective pulse. Each time the outer convective boundary moves outward, it engulfs a new hydrogen-rich shell. The protons are then quickly advected and burnt deeper inside the pulse. This episode is followed by a plateau and readjustment phase during which the top of the pulse does not increase in mass during several models. Then the convective instability increases in mass again, some hydrogen is engulfed and another spike occurs. This discrete ingestion of protons also reflects on the neutron densities as can be seen in the bottom panel of Fig.~\ref{fig_kip2}. Finally, we would like to stress that in the present calculations, we use the strict Schwarzschild criterion to delineate the convective boundaries and do not consider any extra mixing that may give a smoother behavior.\\

\begin{figure}[t]
\includegraphics[width=\columnwidth]{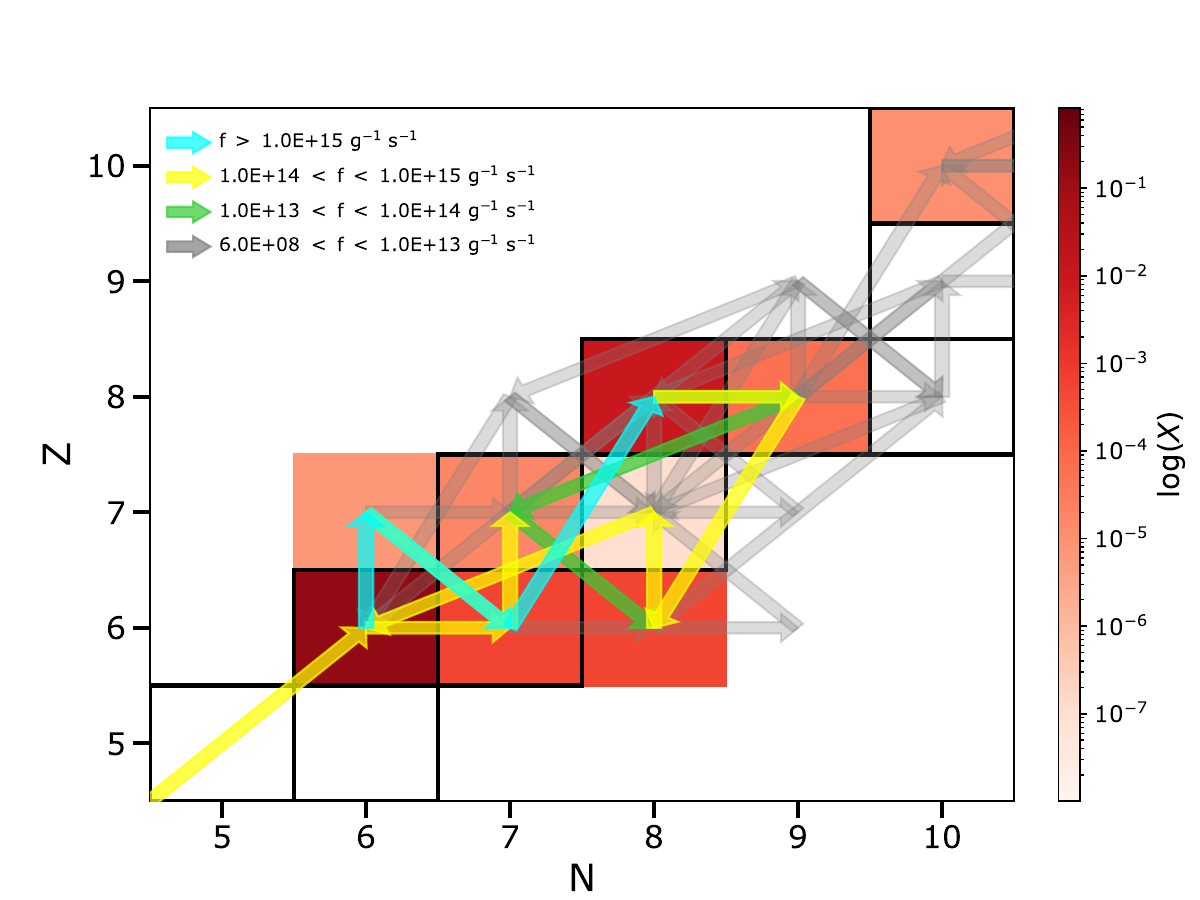}
\caption{Fluxes $f$ during the main proton ingestion event, at the bottom of the convective pulse, when the neutron density is maximal (red point in the top panel of Fig.~\ref{fig_kip2}). 
The fluxes are defined as $f=  N_{\rm av}  \rho Y_i Y_j \langle \sigma v \rangle$ (where $Y_i$ and $Y_j$ are the molar mass fractions of the target and projectile, respectively, $N_{\rm av}$ the Avogadro number and $\langle \sigma v \rangle$ the nuclear reaction rate). 
The colors of the arrow identify various reaction fluxes. The isotopes mass fractions are color-coded according to the right sidebar. The black squares denote stable isotopes. 
}
\label{fig_flow1}
\end{figure}

\subsection{Nucleosynthesis}\label{pulse_nuc}

The main proton ingestion and its associated nucleosynthesis occurs in between the two colored circles in the top panel of Fig.~\ref{fig_kip2}. 
At the time and location corresponding to the red circle, the neutron density reaches its maximum ($4.3 \times 10^{14}$~\cm3; bottom panel of Fig.~\ref{fig_kip2}).
During this event, the chain \iso{12}C(\pg)\iso{13}N($\beta^+$)\iso{13}C(\an)\iso{16}O is very active (Fig.~\ref{fig_flow1}) and provides the great majority of neutrons. 
When neutron density is maximal, the reactions \iso{12}C(\ng)\iso{13}C, \iso{16}O(\ng)\iso{17}O, and \iso{17}O(\na)\iso{14}C are the strongest neutron poisons with fluxes $f > 10^{14}$~g$^{-1}$~s$^{-1}$. 
The reaction \iso{14}N(\np)\iso{14}C is slower with a flux $f = 1.5 \times 10^{10}$~g$^{-1}$~s$^{-1}$.
Despite the strong poisoning effect of some light species such as \iso{16}O, we note that some neutron recycling chains are also at work. For instance, after \iso{16}O(\ng)\iso{17}O and \iso{17}O(\na)\iso{14}C the strong chain \iso{14}C(\pg)\iso{15}N(\pa)\iso{12}C refills $^{12}$C and hence indirectly strengthens the production of neutron via \iso{12}C(\pg)\iso{13}N($\beta^+$)\iso{13}C(\an)\iso{16}O.

\begin{figure}[t]
\includegraphics[width=\columnwidth]{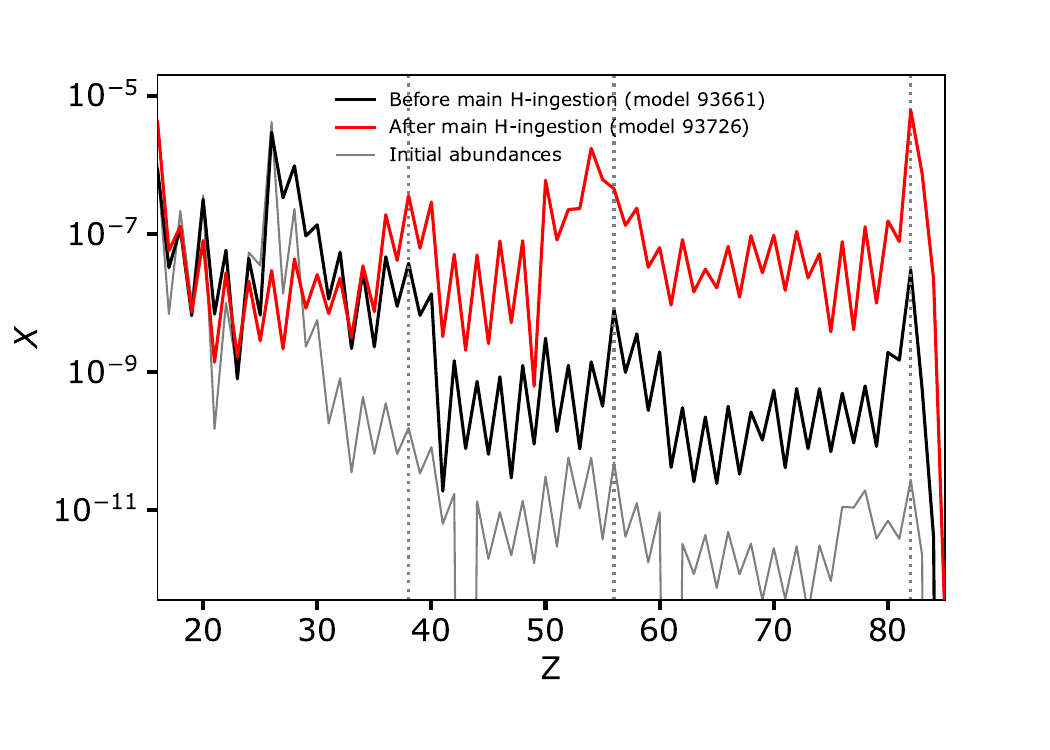}
\caption{Mass fractions of elements just before (black) and after (red) the main proton ingestion, at the bottom of the third convective pulse. These two patterns correspond to the circles of the same color in the top panel of Fig.~\ref{fig_kip2}. The gray line shows the initial abundances of the model. The vertical dashed lines show the location of Sr, Ba, and Pb.
}
\label{fig_ab1}
\end{figure}

The chemical compositions at the bottom of the convective pulse, just before and after the main proton ingestion are shown in Fig.~\ref{fig_ab1}. These two patterns correspond to the two colored circles in the top panel of Fig.~\ref{fig_kip2}. 
As mentioned in Sect.~\ref{strucevol}, before the main proton ingestion event, the stellar model already underwent a weak proton ingestion with neutron densities reaching $N_n = 3 \times 10^{12}$~\cm3, leading to an initial production of trans-iron elements. This explains why the black (before the main ingestion) and gray (initial abundances) patterns in Fig.~\ref{fig_ab1} differ. 
After the splitting, in the intermediate convective zone (at, e.g., $M_{\rm r} = 0.54$~\Msun\ and around model 94000), the chemical composition is similar to the red pattern in Fig.~\ref{fig_ab1}. After the merging with the envelope (around model 94200), the chemical composition is diluted in the envelope but the relative abundance distribution is not affected.

\subsection{$\alpha-$enhancement}\label{alpha_e}

The model discussed in this section was computed with a solar-scaled mixture. As mentioned in \cite{cristallo16}, an $\alpha-$enhanced mixture can have an impact on the proton ingestion episode because of the larger initial oxygen abundance, which boosts the nuclear energy production (through the ON cycle) in the hydrogen-burning shell during the AGB phase. As a test, we computed a similar 1~\Msun{} model with [$\alpha$/Fe]~$=0.5$. We found that the evolution of this model is very similar to that with a solar-scaled composition (especially during the proton ingestion stage).

\subsection{Network size}
\label{sect_netw}

During the main proton ingestion, the simulations were performed with the large network (1091 nuclei). To assess whether the small network represents a good approximation of the full nucleosynthesis, we repeated the same simulation with the small network (411 nuclei). 

\begin{figure}[t]
\includegraphics[scale=0.52, trim = 0cm 1cm 0cm 0cm]{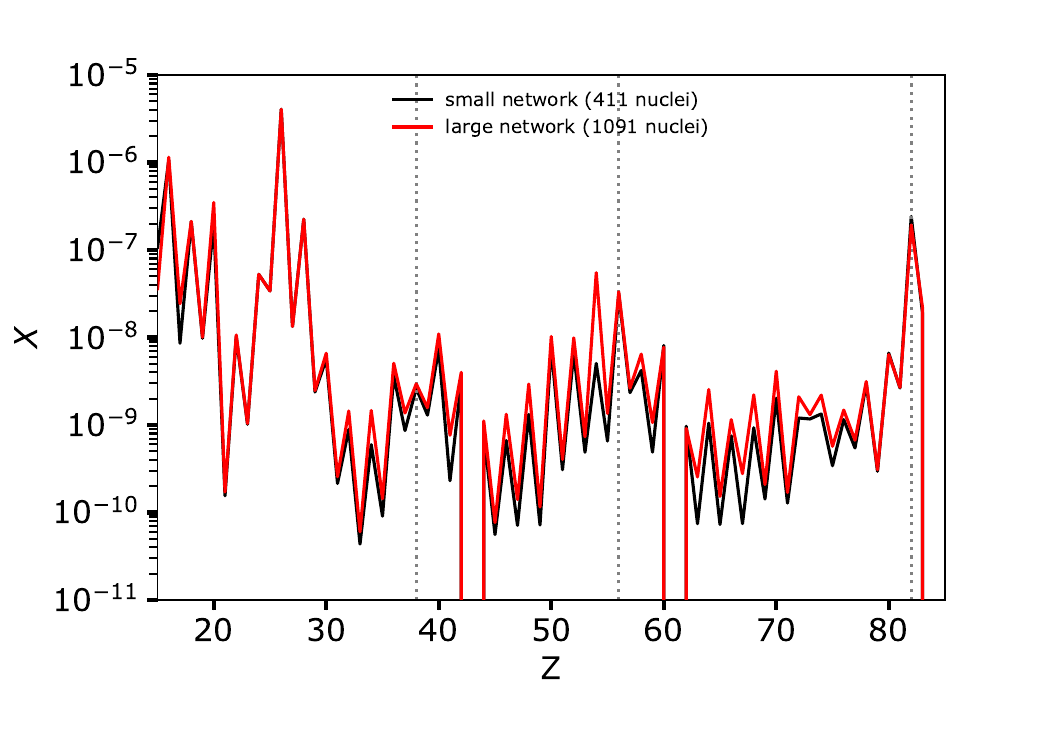}
\includegraphics[scale=0.52, trim = 0cm 0cm 0cm 0cm]{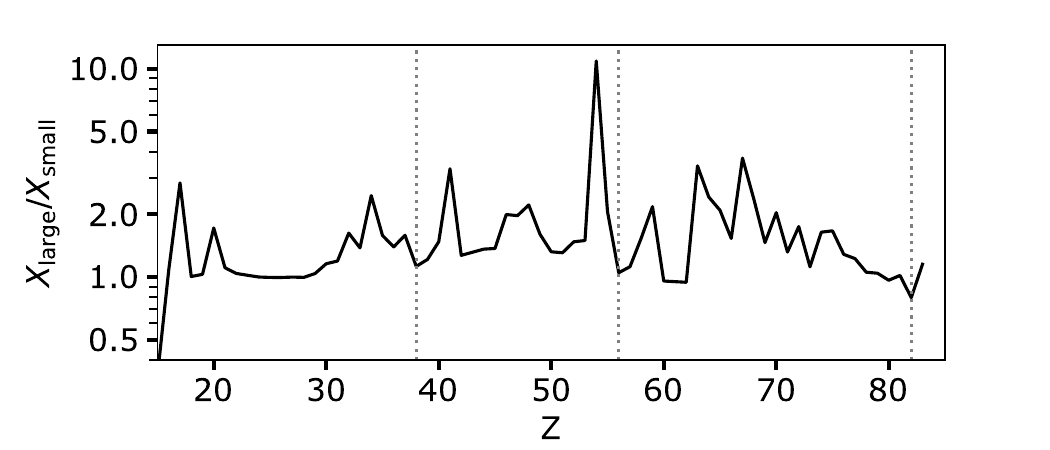}
\includegraphics[scale=0.52, trim = 0cm 0cm 0cm 0cm]{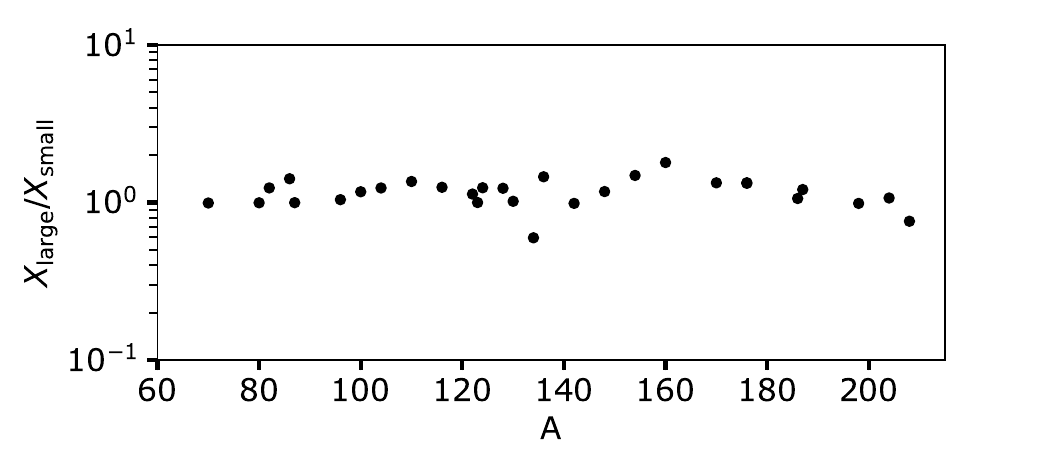}
\caption{Comparison of the surface abundances after the main proton ingestion using the small and large networks (411 and 1091 species, respectively). \textit{Top panel:} Elemental surface mass fractions as a function of the atomic number. \textit{Middle panel:} Ratios of the elemental mass fraction obtained with the small and large networks, as shown on the top panel. \textit{Bottom panel:} Ratio of the isotopic mass fractions obtained with both networks for s-only nuclei as a function of the atomic mass $A$.}
\label{fig_ab_net1}
\end{figure}

As shown in Fig.~\ref{fig_ab_net1}, the small and large networks give rather similar results within a factor of 2--3 except for a few elements. The most noticeable difference concerns Xenon ($Z=54$), which is $\sim 10$ times more produced if the large network is used (Fig.~\ref{fig_ab_net1}, middle panel). 
The main reason for such an impact comes from the rather large production of the stable \iso{136}Xe isotope, which is not included in the small network. Indeed, during a standard s-process with lower neutron densities, the presence of the short-lived \iso{135}Xe (with a half-life of $\sim 9$~h) does not allow the production of \iso{136}Xe. 
However, during the i-process nucleosynthesis, the \iso{135}Xe $\beta$-decay can be bypassed and \iso{136}Xe is produced.

Overall, we note that the production of almost all heavy elements is slightly larger when using the large network. This is balanced by an overproduction of phosphorus ($Z=15$), iron ($Z=26$), and lead ($Z=82$) in the small network model. With the large network, more nuclides can capture neutrons, slowing down the flux toward Pb, which explains its higher abundance in the small network.

For practical reasons, for our sensitivity study (Sect.~\ref{sect_sens}), we consider models computed with the small network, which represents an approximation accurate enough and considerably reduces the computational time. 
In Sect.~\ref{sect_distrib}, for a detailed comparison between s-, i- and r-processes, and in\ Sect. \ref{sect_comp}  for the comparison to observations, we consider models computed with the large network.

\section{Sensitivity study of the nucleosynthetic outputs}
\label{sect_sens}

To ensure that proton ingestion and the associated nucleosynthesis are not numerical artifacts, in addition to our reference simulation (cf. Sect.~\ref{sect_mix}), we computed 11 other models with various spatial and temporal resolutions.

\subsection{Varying the spatial resolution}
\label{subsect_spatial}

At the beginning of each new timestep, the code adds or removes grid points to ensure that the integration variables remain correctly discretized. The addition of a new mesh point is enforced if the variation of a variable $\x$ exceeds a threshold value $\epsilon_\max$ between two adjacent shells. 
In {\sf STAREVOL}, the variables $\x$ are the luminosity, radius, density, temperature, density, and $\ln f$, which in our formulation of the equation of state \citep{pols95} is the new independent variable instead of density or pressure  \citep[$\ln f$ is related to the degeneracy parameter; see][]{siess00}. For a shell $i$, this condition is written as $|1-\x_{i+1}/\x_i|> \epsilon_\max$ where $\epsilon_\max$ typically equals $0.1$. 
Similarly, a shell is deleted if, for all the considered variables,  $|1-\x_{i+1}/\x_i|<  \epsilon_\min$ with $\epsilon_\min$ typically equal to 0.02. We complemented this logic by adding the \iso{1}H and \iso{12}C abundances to the list of variables $\x$.
This check on the chemical profile ensures a correct discretization of the \iso{13}C pocket where the neutrons are produced. 
The mass distribution is also controlled by imposing that between two adjacent shells, the relative mass variation never exceeds $\epsilon_{M,\max}$. Formally this is written as $|dm_i/M| < \epsilon_{M,\max}$,  where $dm_i = m_{i+1}-m_i$ and $\epsilon_{M,\max} \approx 4\times 10^{-3}$. Our reference model was computed with $\epsilon_\max = 0.08$. We also investigated the case $\epsilon_\max = 0.06$ and $0.04$, which can lead to stellar models with more than 3000 shells when reaching the neutron density peak (cf. Table~\ref{table:1}).

\subsection{Varying the temporal resolution}
\label{subsect_temporal}

We also varied the temporal resolution during the proton ingestion. This was done by changing the $\alpha$ parameter defined as 
\begin{equation}
\Delta t^{n+1}  = \alpha \times \Delta t^{n} \times \mathrm{min} \,  \left(\frac{\x_k^n}{| \x_k^n - \x_k^{n-1}|} \right)
,\end{equation}
where $\Delta t^{n}$ is the timestep of the $n^{\rm th}$ model and $\x_k^n$ the values of the structural variables $\ln f$, $T$, $L,$ and $r$ of the $n^{\rm th}$ model at shell $k$. 
The parameter  $\alpha$  controls by how much the timestep can change based on the relative variations of the structure variables between two timesteps. 
The default model was computed with $\alpha = 0.01$. We also investigated the case $\alpha = 0.008$, $0.006,$ and $0.004,$ which results in a reduction of the timestep by a factor of $2-3$ at maximum.

\subsection{Results}
\label{subsect_res}

\begin{figure}[t]
\includegraphics[width=\columnwidth]{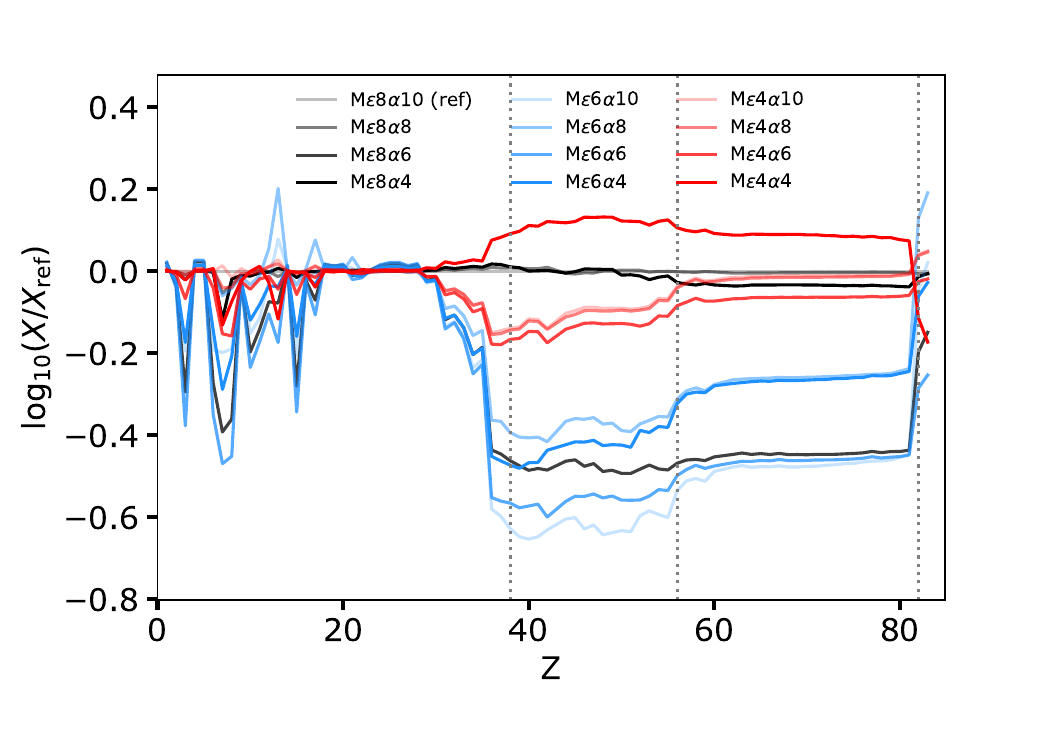}
\caption{Relative variation (in dex) of the surface mass fractions after the main proton ingestion event for models with different spatial and temporal resolutions (cf. Table~\ref{table:1}). The reference model is M$\epsilon 8 \alpha 10$.
}
\label{fig_ab3}
\end{figure}

The properties of our test simulations are summarized in Table~\ref{table:1} and the nucleosynthesis results are compared in Fig.~\ref{fig_ab3} relative to our reference model described in Sect.~\ref{sect_mix} and labeled as M$\epsilon 8 \alpha 10$ in Table~\ref{table:1}. Changing the temporal and spatial resolutions leads to variations in the final surface abundances. Nevertheless, the variation remains modest: compared to the reference model, the surface abundances vary at maximum by $0.6$~dex. No trend with $\epsilon_{\rm max}$ or $\alpha$ really emerges. For instance, we see that the low (black) and high (red) spatial resolution models  give overall similar results (except however for the M$\epsilon$8$\alpha$6 model), while the middle spatial resolution models (blue) predict smaller surface abundances. Similarly, no firm conclusion can be drawn by varying the time resolution for a given $\epsilon_\mathrm{max}$. 
The variations are due to slight structure changes in the models that eventually impact the maximal neutron density (Table~\ref{table:1}), leading to slight differences in the i-process nucleosynthesis.
Also, depending on the model, the spikes due to proton ingestion seen, for example, in the neutron densities (cf. Sect.~\ref{strucevol} and Fig.~\ref{fig_kip2}) are not perfectly identical (especially if the timestep changes). This also affects the nucleosynthesis and eventually the final abundance patterns.
Finally, the splitting of the helium-burning convection zone can occur at slightly different times corresponding to different neutron exposures. 
Consequently, if the splitting does not occur exactly at the same stage of the nucleosynthesis,  differences in the surface chemical patterns can result.

\begin{table}
\scriptsize{
\caption{Characteristics of the models used to assess the impact of temporal and spatial discretization on the resulting nucleosynthesis. 
Columns 2 and 3 are the discretization parameters defined in Sect.~\ref{subsect_spatial} and \ref{subsect_temporal}. Column 4 reports the maximum neutron density and column 5 indicates the mesh number when the neutron density reaches its peak value. Our reference model (described in Sect.~\ref{sect_mix}) is M$\epsilon 8 \alpha 10$.
\label{table:1}
}
\begin{center}
\resizebox{7.cm}{!} {
\begin{tabular}{lcccc} 
\hline
Model   &  $\epsilon_{\rm max}$    & $\alpha$    &  $\log(N_{\rm n,max}$)          &  $N_{\rm shell}$     \\
 label  &                                           &         &                                          &                                    \\
\hline
M$\epsilon 8 \alpha 10$   &  0.08    & 0.01    &  14.64 &   2085     \\
M$\epsilon 8 \alpha 8$   &  0.08    & 0.008    &14.63  &  2085     \\
M$\epsilon 8 \alpha 6$   &  0.08    & 0.006    & 14.63  & 2085     \\
M$\epsilon 8 \alpha 4$   &  0.08    & 0.004    & 14.74  & 2091     \\
\hline
M$\epsilon 6 \alpha 10$   &  0.06    & 0.01    &  14.65 &  2324     \\
M$\epsilon 6 \alpha 8$   &  0.06    & 0.008    & 14.65 &   2321     \\
M$\epsilon 6 \alpha 6$   &  0.06    & 0.006    & 14.49 &  2321     \\
M$\epsilon 6 \alpha 4$   &  0.06    & 0.004    & 14.71 &  2341     \\
\hline
M$\epsilon 4 \alpha 10$   &  0.04    & 0.01    &  14.45 & 3096     \\
M$\epsilon 4 \alpha 8$ &  0.04    & 0.008    & 14.45 &  3104     \\
M$\epsilon 4 \alpha 6$ &  0.04    & 0.006    & 14.41 &  3102     \\
M$\epsilon 4 \alpha 4$ &  0.04    & 0.004    & 14.70  &  3111     \\
\hline 
\end{tabular}
}
\end{center}
}
\end{table}

We find that the main proton ingestion episode resists when increasing the temporal and spatial resolution, meaning that it is not a numerical artifact. However if the nucleosynthesis and diffusion equations are not coupled, the timesteps too large or the spatial resolutions too low, this event can be missed and the evolution proceeds in a standard way. 

Among the various models studied in this work, M$\epsilon 4 \alpha 4$ has the highest spatial and temporal resolution and is therefore expected to provide the most reliable results.
Nevertheless, we cannot discard the other models, and therefore the resulting surface abundance pattern (hence the AGB yields) associated with this nucleosynthetic process is subject to uncertainties due to a numerical noise on the order of 0.6~dex ($\pm 0.3$~dex) that can hardly be avoided. 
This sensitivity study clearly points out the importance of a high-resolution modeling of the ingestion mechanism before drawing any conclusion.

For practical reasons, in the next sections, we consider the model M$\epsilon 8 \alpha 10$ (which is our reference model, discussed in Sect.~\ref{sect_mix}) computed with the large network during proton ingestion, since it was shown to give structural and nucleosynthesis predictions very close to the M$\epsilon 4 \alpha 4$ model (difference of about 0.1~dex at most - see Fig.~\ref{fig_ab3}).

\section{Abundance signatures of the i-process}
\label{sect_distrib}

The chemical abundances resulting from i-process nucleosynthesis are different than those resulting from the r- or s-process.
In this section, we identify the elements and isotopes that can help discriminate between the various neutron capture processes.

\subsection{AGB star s-process versus i-process}
\label{subsec_sproipro}

\begin{figure}[t]
\includegraphics[scale=0.52, trim = 0cm 1cm 0cm 0cm]{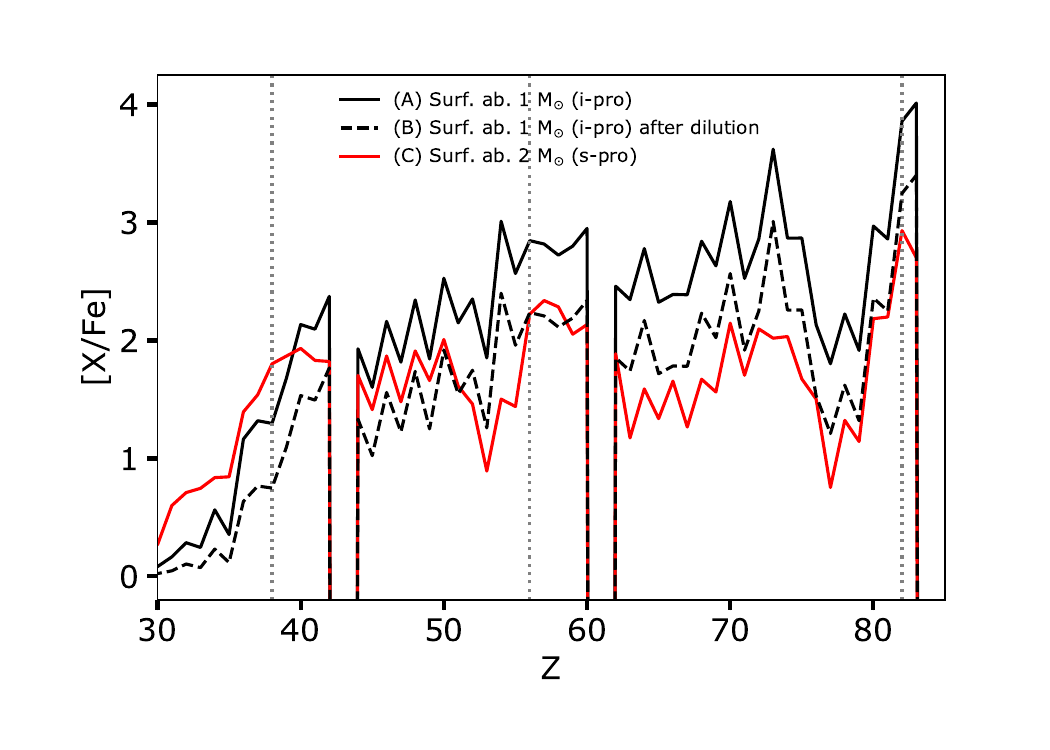}
\includegraphics[scale=0.52, trim = 0cm 0cm 0cm 0cm]{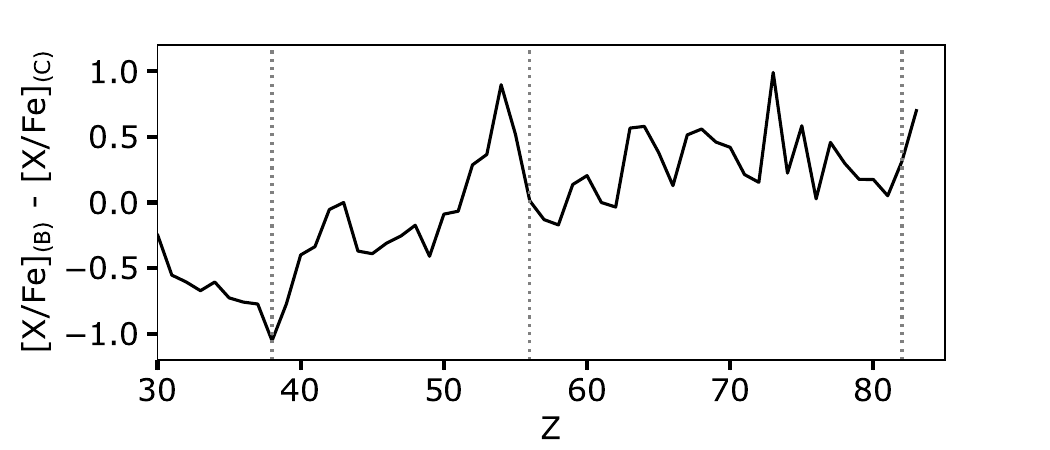}
\caption{Surface enrichment in AGB stars after the development of the s- and i-processes. \textit{Top panel:} Final surface [X/Fe] ratios obtained after the development of the i-process in our reference 1 \Msun\ AGB model and the s-process in a 2 \Msun\ at [Fe/H]~$=-2.5$ (see text for details). 
The model (B) is obtained after dilution of the final surface abundances of model (A) with a solar-scaled abundance distribution to reproduce the Ba (Z = 56) abundance of the model (C).
\textit{Bottom panel:} Difference between model (B) and (C).  
}
\label{fig_sicomp}
\end{figure}

In the top panel of Fig.~\ref{fig_sicomp} we show the final surface overproduction factors [X/Fe] of a model experiencing an i-process and an s-process. The s-process surface abundances are obtained from a  2~$\msun$ model star at [Fe/H]~$=-2.5$ with some overshoot at the base of the convective envelope, as described in \cite{goriely18c} and \cite{karinkuzhi21}; we defer an extended study on the effect of overshoot and initial mass to a future paper. 
After a rescaling of the i-process pattern, significant differences persist with respect to the s-process predictions (bottom panel of Fig.~\ref{fig_sicomp}). In particular, the i-process model generally underproduces elements with $Z \lesssim 50$ by $\sim 0 - 0.5$~dex; 
 the i-process model generally overproduces elements with $Z \gtrsim 50$ by $\sim 0 - 0.5$~dex; the Xe ($Z=54$) and Ta ($Z=73$) are overproduced by about 1 dex in the i-process model.

The two first points result in significant differences in the ratios of light to heavy elements between the two models. For instance, the [hs/ls] ratio (where hs represents the heavy s-elements, such as La, Ba, or Ce, and ls the light s-elements, such as Sr, Y, or Z) are quite different. In particular, the [Ba/Sr] ratios are about 1.2 and 0.4 for the i- and s-process models, respectively.

The overproduction of Xe in the i-process model is due to the large production of \iso{136}Xe, which cannot be produced during s-process (cf. Sect.~\ref{sect_netw}). 
Ta is also found to be significantly overproduced by the i-process model because of the large production of the stable \iso{181}Ta isotope, which comes from the $\beta$-decays of the unstable neutron rich \iso{181}Yb ($Z=70$), \iso{181}Lu ($Z=71$), and \iso{181}Hf ($Z=72$) isotopes. In the s-process, only \iso{181}Hf is produced and decays (with a half life on the order of a few days) to \iso{181}Ta.

\subsection{Isotopic ratios}

Beyond elemental ratios, isotopic ratios can help to discriminate between several neutron capture processes. The isotope fraction $f_{\rm iso}$ of a given species having a number density $n_{\rm iso}$ can be computed as
\begin{equation}
\label{eq_fiso}
 f_{\rm iso} =  \frac{n_{\rm iso}}{\sum_{j=1}^{m} n_{j} }
\end{equation}
where the sum goes over the $m$ stable isotopes of the associated element and where $n_{j}$ is the number density of isotope $j$. For instance, the isotope fraction of \iso{13}C is given by
\begin{equation}
 f_{^{13}C} =  \frac{n({^{13}{\rm C}})}{ n({^{12}{\rm C}}) + n({^{13}{\rm C}})}.
\end{equation}

\begin{figure*}[t]
\includegraphics[scale=0.7, trim = 0cm 1.5cm 0cm 0cm]{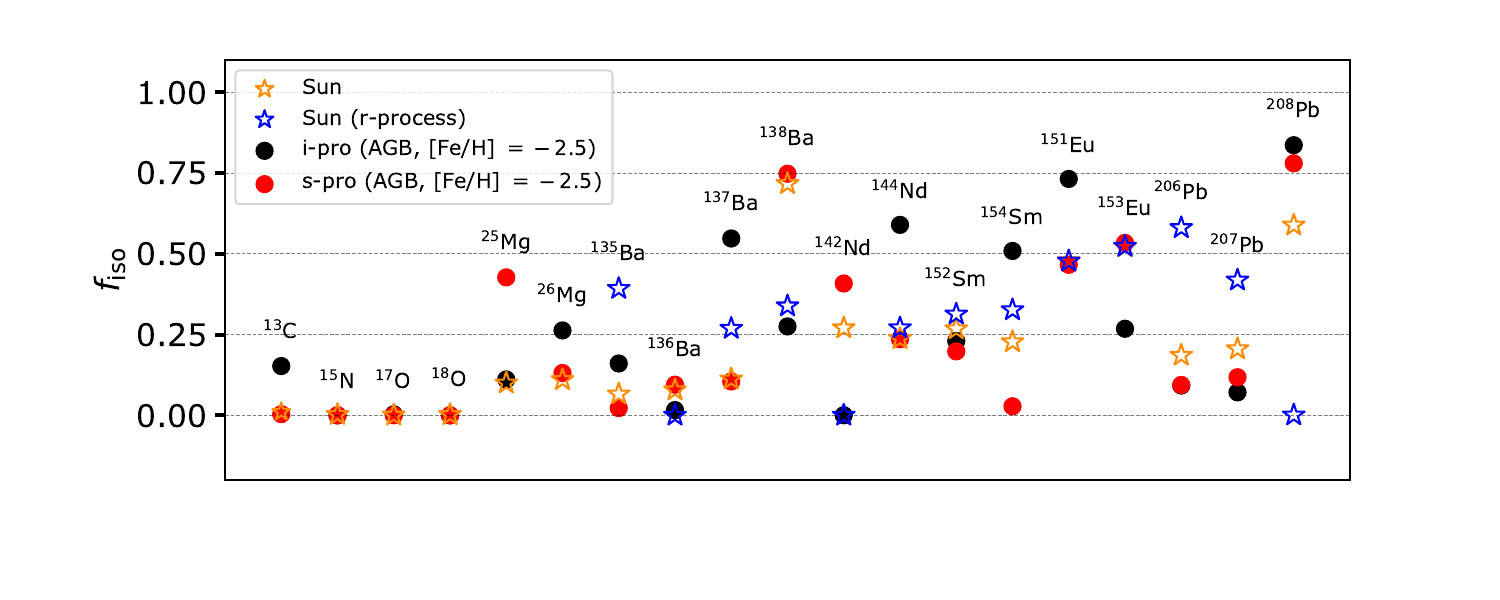}
\caption{Fraction $f_\mathrm{iso}$ (Eq.~\ref{eq_fiso}) of various isotopes for the Sun \citep[orange stars, abundance data from][]{asplund09}, the solar r-process \citep[blue stars, abundance data from][]{arnould07}, our 1 \Msun , [Fe/H]~$=-2.5$ reference AGB model experiencing proton ingestion (black circles, model A in Fig.~\ref{fig_sicomp}), and a 2~\Msun\ AGB model experiencing s-process, at the same metallicity (red circles, model C in Fig.~\ref{fig_sicomp}). We note that $f_{\rm Ba, odd} = f_{\rm ^{135}Ba} +  f_{\rm ^{137}Ba}$.}
\label{fig_iso}
\end{figure*}

\begin{figure}[t]
\centering
\includegraphics[scale=0.6, trim = 0.5cm 0.8cm 0cm 0cm]{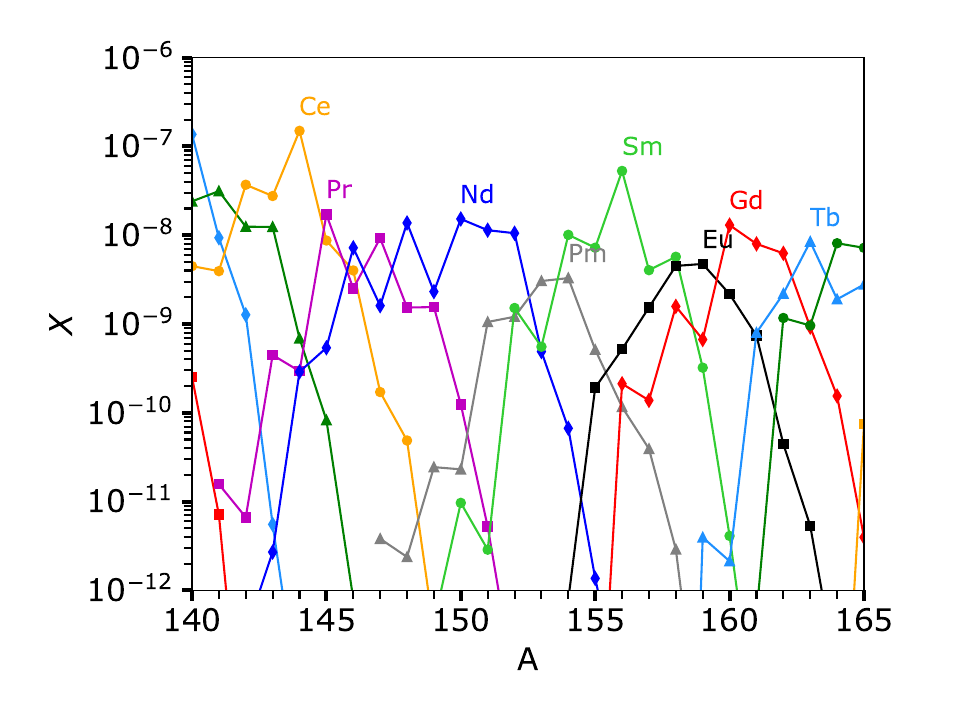}
\caption{
Mass fractions of the isotopic chains as a function of the atomic mass A, just after the peak of neutron density, at the bottom of the convective helium shell (red point in Fig.~\ref{fig_kip2}).
}
\label{isochains}
\end{figure}

We show in Fig.~\ref{fig_iso} that depending on the isotopes, an AGB star experiencing the s- or i-process would sometimes show very different $f_{\rm iso}$ values. We focus below on Ba, Nd, Sm, and Eu, which present interesting features and whose determination from observation is possible, although complicated (see Sect.~\ref{sect_isotobs}).\begin{itemize}
\item In the i-process model, \iso{137}Ba is more abundant than \iso{138}Ba. The \iso{137}Ba isotope is populated by the decay of the neutron-rich \iso{137}Cs isotope, the latter not being produced by the s-process nucleosynthesis. This eventually leads to noticeable differences in the $f_{\rm ^{137}Ba}$ and $f_{\rm ^{138}Ba}$ values for the s- and i-process models.
\item The isotope fractions of Nd also show significant differences. During s-process nucleosynthesis, \iso{142}Nd is the most produced Nd isotope, before \iso{144}Nd and \iso{146}Nd. In the i-process model, the high final  abundance of \iso{144}Nd comes from the decay of \iso{144}Ce, which is largely produced (Fig.~\ref{isochains}) owing to the strong neutron flux.
\item The stable \iso{154}Sm isotope is only produced by the i-process model owing to the short half-life of \iso{153}Sm ($\sim 50$~h), which prevents any production of \iso{154}Sm during a classical s-process nucleosynthesis.
\item The high neutron flux also leads to the production of europium isotopes with masses ranging from $A=155$ to $162$ (Fig.~\ref{isochains}). All these unstable isotopes later decay to the valley of stability. The relatively high final abundance of the stable \iso{151}Eu isotope is due to the large production of the neutron-rich \iso{151}Nd isotope, which after several decays leads to \iso{151}Eu. The stable \iso{153}Eu isotope is populated by the decay of \iso{153}Nd, \iso{153}Pm, and \iso{153}Sm, but does not reach the level of enrichment of \iso{151}Eu. This explains the high $f_{\rm ^{151}Eu}$ and low $f_{\rm ^{153}Eu}$ characterizing the i-process. 
\end{itemize} 

Overall, $f_{\rm ^{137}Ba}$ (or $f_{\rm Ba,odd}${\footnote{$f_{\rm Ba, odd} = f_{\rm ^{135}Ba} +  f_{\rm ^{137}Ba}$.}), $f_{\rm ^{144}Nd}$, $f_{\rm ^{154}Sm}$, $f_{\rm ^{151}Eu}$, and $f_{\rm ^{153}Eu}$ could be very good tracers of i-process nucleosynthesis since they clearly differ from the values of other processes. Moreover, any combination of s- and r-process nucleosynthesis would generally not be able to reproduce the isotopic fractions associated with the i-process nucleosynthesis. For instance, any combination of s- and r-process cannot lead to $f_{\rm ^{151}Eu} > 0.5$ (Fig.~\ref{fig_iso}) while our i-process model predicts $f_{\rm ^{151}Eu} > 0.5$.


\section{Comparison to observations}
\label{sect_comp}

If the AGB model discussed in the previous section exists in nature, it could let its nucleosynthetic imprint on other stars. There is a growing sample of metal-poor stars showing abundance patterns midway between the s- and r-processes. These r/s-stars are good candidates to host the nucleosynthetic signature of the AGB models discussed above.

\subsection{Selection of sample of r/s-stars}
\label{subsect_selec}

To determine the sample of r/s-stars, we follow \cite{karinkuzhi21} who propose a new criterion for classifying s-, r-, and r/s-stars. 
This criterion is based on the average abundance distance $d_{\rm RMS}$ to the solar scaled r-process, which can be expressed as
\begin{equation}
d_{\rm RMS} = \left( \frac{1}{N} \sum_{i=1}^{N} (A_{ i,\star} - A_{i,r})^2 \right)^{1/2}.
\label{drms}
\end{equation}
In Eq.~\ref{drms}, $A_{ i,\star} = \log_{10}(n_{i,\star}/n_{H,\star}) + 12$ with $n_{i,\star}$ the number density of element $i$ in the sample star.  The quantity $A_{i,r}$ is the solar r-process abundance scaled to the Eu abundance of the sample star.\ This abundance is computed as
\begin{equation}
A_{i,r} = A_{i,r,\odot} - (A_{\mathrm{Eu},r,\odot} - A_{\mathrm{Eu},\star})
\label{Air}
,\end{equation}
where $A_{i,r,\odot}$ is the solar r-process abundance of element $i$ taken from \cite{arnould07}. The $d_{\rm RMS}$ value gives an estimate on how far a sample star is to the solar r-process. It has the advantage of including more than just one or two abundances  for classifying stars (typically [Ba/Fe] and [Eu/Fe] for r/s-stars).
As shown in Figure 10 of \cite{karinkuzhi21}, the s-, r-, and r/s-stars are nicely separated when using this criterion. 
The r-stars (s-stars) have the lowest (highest) $d_{\rm RMS}$ values.
The r/s-stars are characterized by intermediate $d_{\rm RMS}$ values: $0.5 \leq d_{\rm RMS} \leq 0.8$.

\cite{karinkuzhi21} apply this new classification scheme to a sample of 25 stars (with $-3.35<$~[Fe/H]~$<-0.98$) observed with the high-resolution HERMES spectrograph. They find that only two objects in their sample are classified as s-stars with the new method, while classified as r/s-stars with the original method. In the end these authors found 13 s-stars, 11 r/s-stars, and 1 r-star.

In this work, we extend the \cite{karinkuzhi21} study to all known r/s-stars. We first selected the stars in the SAGA database\footnote{We also included the stars of \cite{karinkuzhi21}, which are not yet included in the SAGA database.} \citep{suda08} having 
\begin{itemize}
\item a metallicity of $-2.7 <$~[Fe/H]~$<-2.3$ (i.e., [Fe/H]~$=-2.5\pm 0.2$),
\item an enrichment in barium relatively to iron [Ba/Fe] $> 0.5$,
\item at least five measured abundances between Ga (Z=31) and Pb (Z=82).
\end{itemize}
This gives us a sample of 43 stars with $d_{\rm RMS}$ values ranging between $0.18$ and $1.48$. 
We then applied the \cite{karinkuzhi21} criterion ($0.5 \leq d_{\rm RMS} \leq 0.8$) to obtain a sample of 14 stars, whose characteristics are reported in Table~\ref{table_fits}. We note that in our metallicity range, 6 out of these 14 stars are in the sample of \cite{karinkuzhi21}.

\begin{figure*}[t]
\centering
\includegraphics[scale=0.85, trim = 3cm 0.0cm 2cm 0cm]{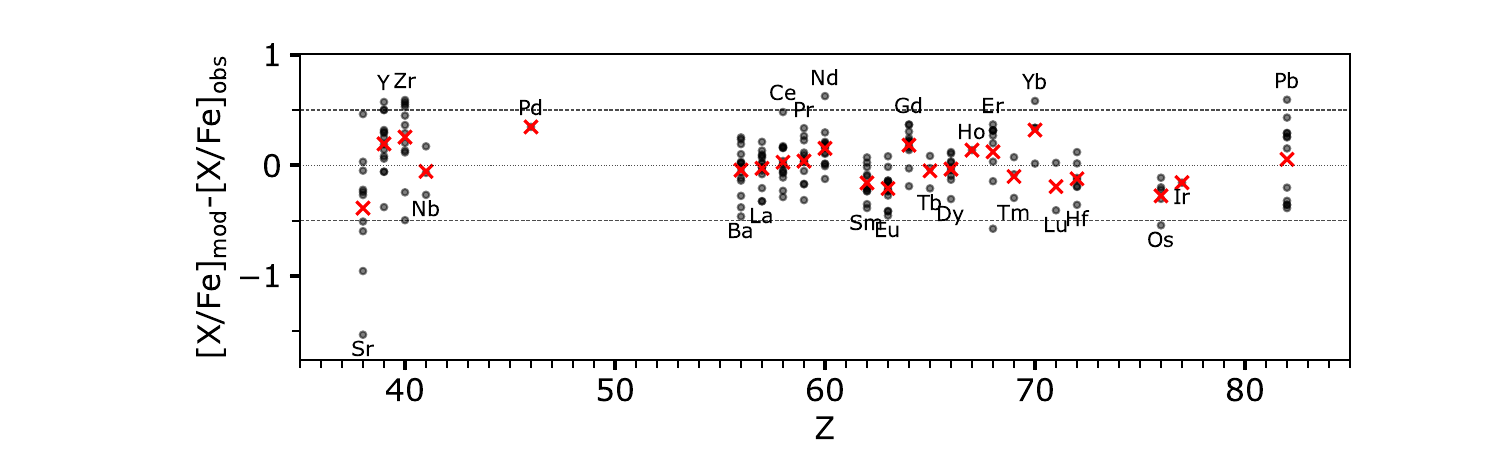}
\caption{Residual of the best fits for the 14 r/s-stars of Table~\ref{table_fits} using our reference 1 \Msun, [Fe/H]~$=-2.5$ AGB model discussed in this paper that experiences a proton ingestion. The red crosses represent the average values of the residuals. The individual fits are shown in Fig.~\ref{fig_fits}.
}
\label{fig_res}
\end{figure*}

\subsection{Fitting recipe}

For a given sample star with $N$ data points and $U$ upper limits, the reduced $\chi_{\nu}^2$ can be computed as \citep[see, e.g.,][]{heger10,ishigaki18,choplin19}
\begin{equation}
\chi^{2}_{\nu} = \frac{1}{N}\left( \sum^{N}_{i=1} \frac{(M_X - C_X)^2}{\sigma^2_{\rm X} }
                +\sum^{N+U}_{i=N+1} \frac{(M_X - C_X)^2}{\sigma^2_{\rm X} } \Theta(M_X - C_X) \right)
\label{eq_chi}
,\end{equation}
where $C_{X}$ and $M_{X}$ are the [X/Fe] ratios derived from observations and predicted by AGB models, respectively; $\Theta$ is the Heaviside function; and $\sigma_{\rm X}$ is the observational uncertainty associated with element $X$. We set a minimal uncertainty of 0.2 (i.e., if $\sigma_{\rm X}<0.2$ we set $\sigma_{\rm X} = 0.2$). 
The second term in Eq.~\ref{eq_chi} allows a proper computation of the $\chi^{2}_{\nu}$ when upper limits exist in the sample. 
If $M_{X} > C_{X}$, then $\Theta = 1,$ and the difference between the fit and the upper limit value is included in the $\chi^{2}_{\nu}$. 
In contrast, if $M_{X} < C_{X}$, then $\Theta = 0,$ and the difference between the fit and the upper limit value is discarded. The $1/N$ factor is added to obtain the reduced $\chi^{2}$, which gives an estimate of the fit quality by taking into account the number of data points $N$. 
To calculate the $\chi^{2}_{\nu}$, we considered all available elements heavier than Zn (i.e., with $Z>30$). 

For each sample r/s-star, the minimum $\chi^{2}_{\nu}$ is searched using the final surface abundances (after decays) of our reference AGB model diluted with some initial material (i.e., the interstellar material with which the star formed). 
The dilution factor is freely varied so as to minimize the $\chi^{2}_{\nu}$ value. 
The mass fraction $X_{i}$ of an isotope $i$, diluted with some ISM material is calculated according to the relation
\begin{equation}
X_{i} =  (1-f) \, X_{s} + f \, X_{\rm ini}
\label{eq_abdil}
,\end{equation}
where $X_{s}$ and $X_{\rm ini}$ are the surface and initial mass fractions of isotope $i$, respectively, and $0 \leq f \leq 1$ is the dilution factor. 
The [X/Fe] ratio of an element $k$ is computed as follows:
\begin{equation}
[\textrm{X/Fe}] = \log_{10}(Y_{ k} / Y_{\rm Fe}) - \log_{10}(Y_{ k,\odot} / Y_{\rm Fe,\odot})
\label{yie}
,\end{equation}
where $ Y_{ k} = \sum_{j=1}^{m} X_{ j} / A_{ j} $ is the elemental molar fraction of the element $k$, which has $m$ isotopes with mass fractions $X_{j}$ and atomic masses $A_{ j}$.

For comparison, we also fitted the individual abundance distributions of our sample stars with the s-process abundances of our 2~$\msun{}$, [Fe/H]~$=-2.5$ AGB model (cf. Sect.~\ref{subsec_sproipro}), following the same procedure as described above. Finally, we repeated the same procedure using the solar r-process \citep[from][]{arnould07}. In this case, the solar r-process pattern is just scaled so as to minimize the $\chi_{\nu}^2$ value.

  \begin{table}
\scriptsize{ 
\caption{Characteristics and fit parameters of the 14 selected r/s-stars. The quantity $N$ is the number of data points between Zn ($Z=30$) and Bi ($Z=83$), $d_{\rm RMS}$ the average distance to the solar scaled r-process (Eq.~\ref{drms}), $\chi_{\rm \nu,min}^2$ the minimum reduced $\chi_{\nu}^2$ value (Eq.~\ref{eq_chi}), and $f$ the dilution factor (Eq.~\ref{eq_abdil}).}
\label{table_fits}
\begin{center}
\resizebox{9cm}{!} {
  \begin{tabular}{lllllll}
\hline
 Star        & [Fe/H] & $N$ & $d_{\rm RMS}$ & $\chi_{\rm \nu,min}^2$ & $f$  & Ref                \\
\hline
 CS22891-171 & -2.5   & 14           & 0.79           & 1.19            & 0.86 & 1, 2              \\
 CS22948-027 & -2.26  & 10           & 0.60           & 1.56            & 0.78 & 3, 4, 5           \\
 CS29497-030 & -2.52  & 13           & 0.77           & 2.24            & 0.72 & 6, 7, 8           \\
 CS31062-050 & -2.48  & 23           & 0.76           & 1.85            & 0.82 & 9, 10, 11, 12, 13 \\
 HD145777    & -2.32  & 13           & 0.79           & 2.94            & 0.98 & 1                 \\
 HD187861    & -2.6   & 15           & 0.70           & 2.18            & 0.88 & 1, 2              \\
 HD224959    & -2.36  & 16           & 0.80           & 5.58            & 0.77 & 1, 2              \\
 HE0243-3044 & -2.58  & 14           & 0.79           & 2.3            & 0.75 & 14                \\
 HE0338-3945 & -2.43  & 17           & 0.76           & 1.2             & 0.73 & 15                \\
 HE1405-0822 & -2.37  & 19           & 0.62           & 1.56            & 0.96 & 16                \\
 HE1429-0551 & -2.7   & 10           & 0.70           & 0.95            & 0.95 & 1                 \\
 HE2122-4707 & -2.47  & 6            & 0.57           & 0.92            & 0.8 & 17                \\
 HE2148-1247 & -2.35  & 13           & 0.80           & 0.56            & 0.68 & 17, 18            \\
 HE2258-6358 & -2.67  & 18           & 0.75           & 1.40            & 0.87 & 19                \\
\hline
\end{tabular}
}
\end{center}
}
\textbf{References}. 1 - \cite{karinkuzhi21}; 2 - \cite{masseron10}; 3 - \cite{aoki02}; 4 - \cite{aoki07}; 5 - \cite{barbuy05}; 6 - \cite{roederer14a}; 7 - \cite{ivans05}; 8 - \cite{sivarani04}; 9 - \cite{lai07}; 10 - \cite{aoki02}; 11 - \cite{lai04}; 12 - \cite{johnson04}; 13 - \cite{aoki06}; 14 - \cite{hansen15}; 15 - \cite{jonsell06}; 16 - \cite{cui13}; 17 - \cite{cohen13}; 18 - \cite{cohen03}; 19 - \cite{placco13}.
\end{table}

\subsection{Fitting results}

The fit residuals and $\chi_{\nu}^2$ values of the best fits are shown in Figs.~\ref{fig_res} and \ref{fig_chi}, respectively (individual fits are presented in Fig.~\ref{fig_fits}). Table~\ref{table_fits} reports the fit parameters of the 14 stars. Fig.~\ref{fig_chi} also shows the results obtained with our s-process model and with the solar scaled r-distribution.

Most of the stars can be reasonably well fitted. Most of the residuals stay in the range $\pm 0.5$~dex (Fig.~\ref{fig_res}) and most of the average residual are in the range $\pm 0.2$~dex. A scatter of $\sim \pm 0.5$~dex in Fig.~\ref{fig_res} is acceptable knowing that observations have typical uncertainties of  $0.2 - 0.4$~dex and that our model suffers a numerical noise inducing an uncertainty of about $\pm 0.3$~dex on the abundances (cf.~Sect.~\ref{subsect_res}).

Figure~\ref{fig_chi} clearly shows that our sample is much better reproduced with an i-process model than with a s- or a r-process model. Only the abundances of HD145777 are better accounted for by the s-process model. Nevertheless, as discussed in the next section, none of the models are able to account for the strong element-to-element variation around $Z=40$ in this star.

\subsubsection{Problematic stars}
\label{sect_probstars}

There are a few problematic stars, with strong abundance variations between elements of close atomic numbers. For instance, HD224959, which has the highest $\chi_{\nu}^2$ value, has [Sr/Fe]~$=2.26$ and [Y/Fe]~$= 0.5$, where the atomic numbers of Sr and Y are 38 and 39, respectively. Such a gap between two consecutive elements is very hard to reproduce with any of the known neutron capture processes. The high [Sr/Fe] value is largely responsible for the high $\chi_{\nu}^2$. Our best-fit model underestimates [Sr/Fe] by about 1.5~dex. If Sr is removed from the fit,  $\chi_{\nu}^2 = 1.82$ instead of 5.58.

Similarly, the stars HD145777,  CS29497-030, and CS31062-050 show strong abundance variations among Sr, Y, and Zr (Z = 38, 39, 40 respectively). In particular, they have [Sr/Y]~$\simeq 1$, which probably cannot be accounted for by any known neutron capture process. These strong variations among the light s-elements contribute to deteriorate the $\chi_{\nu}^2$ (although less dramatically than for HD224959).

Finally, we mention the dysprosium ($Z=66$) abundance of HE0243-3044, which is about 1 dex lower than the erbium ($Z=68$) abundance. Such a variation between these two close elements cannot be reproduced in this work and tends to increase the $\chi_{\nu}^2$ value.

We also note that the derivation of the abundances from spectroscopic observation can suffer from large uncertainties as a result of the NLTE and 3D effects \citep[e.g.,][and references therein]{asplund09}. For instance, \cite{velichko10} reported positive abundance corrections reaching to 0.33 dex for zirconium ($Z=40$) at solar metallicity, and \cite{bergemann12} found that NLTE corrections for strontium can amount to 0.5~dex in extreme cases.

\begin{figure}[t]
\centering
\includegraphics[scale=0.6, trim = 0cm 0.5cm 0cm 0cm]{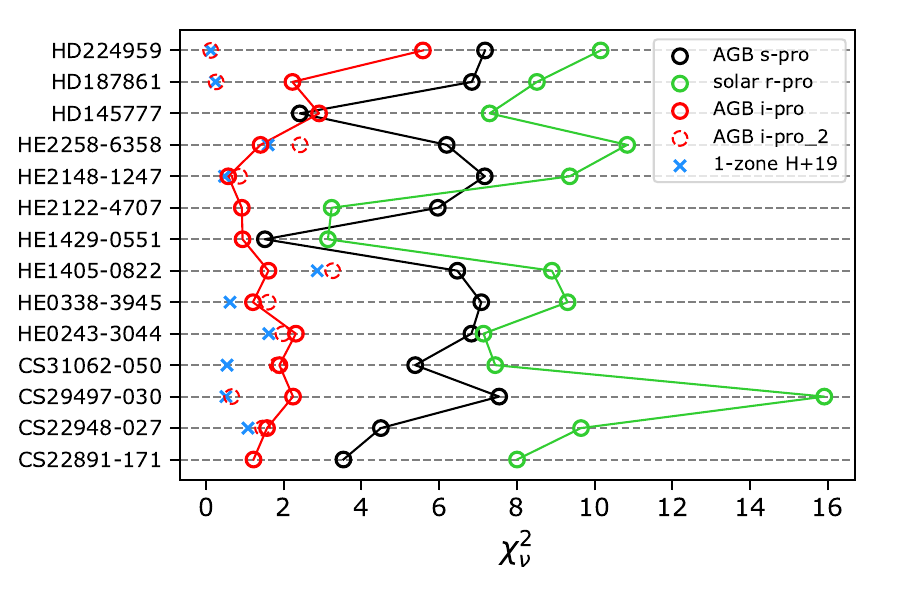}
\caption{
Reduced $\chi$-square values $\chi_{\nu}^2$ of the 14 selected r/s-stars. 
The red, black, and green patterns show the minimum $\chi_{\nu}^2$ values obtained with our reference i-process AGB model, a 2~$\msun{}$ [Fe/H]~$=-2.5$ AGB model experiencing s-process and the solar scaled r-process from \cite{arnould07}, respectively (see text for details).
The blue crosses show the $\chi_{\nu}^2$ values obtained with the one-zone model of \cite{hampel19}. The dashed red circles show the results obtained with our reference i-process AGB model, but with the same abundance data and uncertainties as in \citet[][see text for details]{hampel19}.
}
\label{fig_chi}
\end{figure}

\subsubsection{Comparison with \cite{hampel19}}

\citet[][hereafter H+19]{hampel19} also carried out a $\chi^2$ analysis of r/s-stars by fitting their abundances with the nucleosynthetic outputs of one-zone models computed at constant temperature, density, and neutron density. These authors investigated stars in the metallicity range $-2.8<$~[Fe/H]~$<-1.8$ having [C/Fe]~$>1$, [Eu/Fe]~$>1,$ and [Ba/Eu]~$>0$.
In their sample of 24 stars, 10 are included in our sample. Most of the other stars in \cite{hampel19} are out of our metallicity range ($-2.7<$~[Fe/H]~$<-2.3$) and a few are not included in our sample because they do not satisfy our r/s criterion ($0.5 \leq d_{\rm RMS} \leq 0.8$, cf. Sect.~\ref{subsect_selec}).

The results obtained in this work and H+19 are similar except for a few objects (Fig.~\ref{fig_chi}).
As we show next, these} differences can largely be explained by the different abundance data used to make the fits. 
For instance, while we used 16 data points to fit HD224959 \citep[we used the new abundances of][]{karinkuzhi21}, only 9 were used in H+19, which is less constraining and would tend to give smaller $\chi^2$ values. 
Also, the abundances (and the associated uncertainties) are not always the same, which can lead to significant differences in the $\chi_{\nu}^2$ values.
To address this point, we again fitted the abundances of the stars we have in common in our samples by using exactly the same input abundance data as in H+19. When doing so, the $\chi_{\nu}^2$ value becomes very similar (dashed red circles in Fig.~\ref{fig_chi}). The residual differences now only reflect the (rather small) differences between our simulations and the H+19 models.

\subsection{Isotopic ratios in r/s-stars}
\label{sect_isotobs}

Various studies relate the measurement of isotopic ratios in metal-poor stars. The most discussed ratio is  \iso{12}C/\iso{13}C \citep[e.g.,][]{aoki02, spite06, masseron10, hansen16c}, but barium isotopic fractions \citep[e.g.,][]{magain95,gallagher12,wenyuan18}, as well as europium \citep{sneden02,aoki03a,aoki03b,roederer08}, samarium \citep{lundqvist07,zhang10,roederer08}, and neodymium \citep{roederer08}  were also
largely discussed.

As a word of caution, we note that the derivation of isotopic ratios from observation is very challenging and subject to large uncertainties. For a same star, different studies can lead to significantly different values. For instance, for the metal-poor subgiant HD140283, \cite{magain95} first derived a $f_{\rm Ba, odd}$ ($f_{\rm Ba, odd} = f_{\rm ^{135}Ba} +  f_{\rm ^{137}Ba}$) value of $0.06 \pm 0.06$, but \cite{lambert02} later found $f_{\rm Ba, odd} = 0.30 \pm 0.21$. This was confirmed by \cite{collet09} who estimated $f_{\rm Ba, odd} = 0.33 \pm 0.13$. However, later on, \cite{gallagher10} found $f_{\rm Ba, odd} = 0.02 \pm 0.06$, which is close to the original value predicted by \cite{magain95}.

\subsubsection*{\iso{151}Eu in CS31062-050}

Interestingly, \cite{aoki03b} derived $f_{\rm ^{151}Eu}$ for CS31062-050, which is included in our sample. 
These authors reported $f_{\rm ^{151}Eu}$ values ranging between $0.55 \pm 0.05$ and $0.65 \pm 0.14,$ depending on the atomic line considered to make the analysis (their Table 2). 
Our models predict $f_{\rm ^{151}Eu} = 0.73$ (see also Fig.~\ref{fig_iso}), which is consistent with the \cite{aoki03b} estimate.

\subsubsection*{\iso{135}Ba and \iso{137}Ba in HE0338-3945}

\cite{meng16} derive an $f_{\rm Ba, odd}$ value of $0.23 \pm 0.12$ for the r/s-star HE0338-3945 that is also included in our sample (Table~\ref{table_fits}).
Our model predicts $f_{\rm Ba, odd} = 0.71,$ which is not consistent with the \cite{meng16} observation. 
The r-process predicts $f_{\rm Ba, odd} > 0.5$, hence also too high for this star.
The s-process model is in better agreement with $f_{\rm Ba, odd}$ value of 0.12, but still hardly compatible with the \cite{meng16} measurement. \cite{meng16} suggest that HE0338-3945 formed with a mixture of s-process and r-process. Another interesting possibility that would require just one astrophysical source is that HE0338-3945 formed from an AGB star that experienced both the s- and the i-process. It may be that depending on the mass and metallicity, the proton ingestion occurs later during the AGB phase \citep[e.g., as in][]{herwig11}. In this case, the s-process could have operated before the i-process and a mixture of s- and i-process material may result. It is also conceivable that in more massive stars, after the proton ingestion event, the star resumes a standard TP-AGB evolution, where radiative s-process nucleosynthesis followed by third dredge-ups enrich the envelope in s-elements. This will be explored when investigating AGB models with different mass, metallicity, and mixing assumptions.

\subsubsection*{Barium isotopic ratios with metallicity}

We also mention the study of \cite{mashonkina06}, which derived $f_{\rm Ba, odd}$ for 25 stars with $-1.35 <$~[Fe/H]~$<0.25$. These authors found that $f_{\rm Ba, odd}$ grows from $\sim 0.1$ to $\sim 0.5$ toward lower metallicities (their Figure 6). 
The $f_{\rm Ba, odd}$ value in the solar r-process (s-process) being $\sim 0.7$ ($\sim 0.1$), this result may point toward a higher contribution of r-process at lower metallicity, as they mentioned. However, this would be also consistent with a higher contribution of i-process at lower metallicity because i-process nucleosynthesis also predicts large $f_{\rm Ba, odd}$ values.

\subsection{Accretion on the secondary and dilution factors}

It appears that most of the r/s-stars are in binary systems. For instance, \cite{karinkuzhi21} report that 9 out of their 11 r/s-stars have been detected as binaries. 
A possible scenario \citep[e.g.,][]{abate13} considers that the r/s-stars were born in a binary system as $\sim$~0.8~\Msun\ normal stars with a more massive companion (e.g., 1.0~\Msun). During its AGB stage, the massive companion produces and ejects, through winds, a material enriched in trans-iron elements. Some of the wind material is accreted by the lower-mass star that becomes enriched in r/s elements while the companion evolves toward the white dwarf stage.

In this scenario, the dilution factor $f$ (Eq.~\ref{eq_abdil}), which was freely varied to minimize the $\chi_{\nu}^2$ value, can also be linked to physical quantities related to the binary system properties. 
Introducing the accretion efficiency parameter $\beta$, the dilution factor can be written as

\begin{equation}
f = \frac{M_{\rm env}}{M_{\rm env} + M_{\rm acc}} = \frac{M_{\rm env}}{M_{\rm env} + \beta M_{\rm wind}}
\label{dilfac2}
,\end{equation}
where $M_{\rm env}$ is the envelope mass of the r/s-star before the accretion episode, $M_{\rm acc}$ is the mass accreted by the r/s-star, and $M_{\rm wind}$ is the mass lost by the AGB companion through winds. 
In our case, we have $M_{\rm wind} \simeq 0.5$~\Msun\ (Fig.~\ref{fig_kip1}) and $0.68< f <0.98$ (Table~\ref{table_fits}). 
Stellar models of 0.8~\Msun\ at [Fe/H]~$=-2.5$ predict an envelope mass of $M_{\rm env} \simeq 5 \times 10^{-4}$~\Msun\ on the main sequence and of $M_{\rm env} \simeq 0.375$~\Msun\ in the giant phase at the deepest extent of the envelope during the first dredge-up.

There are three main-sequence r/s-stars in our sample: CS29497-030, HE0338-3945, and HE2148-1247. For these stars, Eq.~\ref{dilfac2} gives an accretion efficiency on the order of $\beta \simeq 5 \times 10^{-4}$.
This is somewhat lower than predictions from hydrodynamic calculations\footnote{We note that, in these works, the accretion efficiency is defined as $\beta = \dot{M}_{\rm acc} / \dot{M}_{\rm wind}$. In our simplistic approach, we suppose that $\beta = {M}_{\rm acc} / {M}_{\rm wind}$, which is equivalent of taking the averaged values of $\dot{M}_{\rm acc}$ and $\dot{M}_{\rm wind}$.} \citep{nagae04, liu17, saladino18, saladino19}. 
This difference can be explained by the action of transport processes such as thermohaline mixing \citep[e.g.,][]{eggleton2006,charbonnel07, stancliffe07, stancliffe08}. It is likely that the higher mean molecular weight of the accreted matter allows the sinking of this material in the stellar interior. \cite{stancliffe08} showed that thermohaline instabilities could potentially mix the accreted material over $88\%$ in mass of their 0.8~\Msun\ models. In the case of efficient thermohaline mixing, our predicted $\beta$ value would therefore increase by a factor of about 10, which would better match the predictions from hydrodynamic models.

The other 11 giant r/s-stars in our sample have $0.75 < f < 0.98$. With $M_{\rm env} \simeq 0.375$~\Msun{}, we find that the amount of mass that must be accreted, $M_{\rm acc}$, to account for the enrichment, ranges between $8 \times 10^{-3}$~$M_{\odot}$ and $0.125$~$M_{\odot}$ with a mean of 0.06~$M_{\odot}$. 
These values are realistic and are associated to an accretion efficiency parameter, $\beta$ (Eq.~\ref{dilfac2}), ranging between 0.015 and 0.25. 
This is compatible with 3D hydrodynamical simulations of wind mass transfer in AGB binary systems \citep[e.g. Table 3 of][]{saladino19}.


\section{Summary and conclusions}
\label{sect_conc}

Using a nuclear network of 1091 species coupled to the chemical transport equations, we investigated with the {\sf STAREVOL} code  the development of the i-process in a high-resolution 1~\Msun\,, [Fe/H]~$=-2.5$ stellar model, caused by a proton ingestion event during the early AGB phase. 

We found that, after a minor proton ingestion event during the second thermal pulse ($N_n \sim 3 \times 10^{12}$~\cm3), a second and main proton ingestion event occurs during the third pulse. This event leads to a maximal neutron density of $\sim 4.3 \times 10^{14}$~\cm3 at the bottom of the helium driven convective zone, and triggers an efficient i-process nucleosynthesis. 
Contrary to the first minor proton ingestion event, the second episode (which last for approximately one year) deeply affects the subsequent AGB evolution. The pollution of the envelope and in particular the enhancement of carbon raises the  surface opacity, which leads to an expansion of the star and large increase in the mass-loss rate. The convective envelope is quickly lost, preventing the occurrence of any further thermal pulses.

We found that proton ingestion events resist under various numerical and temporal resolutions, discarding the possibility of a numerical artifact. 
However, by changing the resolution, we found that the resulting abundances of trans-iron elements are subject to a numerical uncertainty of about $\pm 0.3$~dex at maximum on the abundances. 
Our sensitivity study also stresses the need for high spatial and temporal resolutions and the coupling of diffusion and nucleosynthesis equations in these types of calculations. If not, the nucleosynthesis predictions may not be correct or the proton ingestion episode may even be missed.

We then highlighted the nucleosynthetic signatures of this process compared to an AGB s-process and a solar r-process. Beyond the global overproduction of heavy to light trans-iron elements and the fact that our i-process model overproduces Xe and Ta by $\sim 1$ dex compared to the s-process, we found that the isotopic fractions $f_\mathrm{^{137}Ba}$, $f_{\rm ^{144}Nd}$, $f_{\rm ^{154}Sm}$, $f_{\rm ^{151}Eu}$, and $f_{\rm ^{153}Eu}$ are good tracers of i-process nucleosynthesis since they largely differ from the values of the s- and r-processes. 
Our conclusions are nevertheless limited to a mass of 1~\Msun\ and metallicity of [Fe/H]~$=-2.5$. An exploration of the parameter space remains to be done.

The comparison of our i-process AGB model with 14 r/s-stars selected with the new method of \cite{karinkuzhi21} suggests that our model provides a good explanation for these stars, although some discrepancies remain in a few objects because of strong abundance variations between elements of close atomic numbers (especially Sr, Y, and Zr). 

As already suggested in previous works, our study shows that an i-process can naturally occur during the AGB phase of low-metallicity AGB stars. 
This however does not exclude other astrophysical sites for the i-process \citep[e.g.,][]{denissenkov17, clarkson18}.
Regarding the i-process in AGB stars, among the questions that need to be addressed, for instance, are how the i-process depends on the AGB initial mass, metallicity, and mixing (e.g., overshoot, rotation) and how nuclear uncertainties affects the resulting i-process abundances. We will address these points in future studies.

\section*{Acknowledgments}
We are grateful to the the anonymous referee for very insightful comments that contributed to improve the manuscript. 
We thank S. Van Eck, D. Karinkuzhi and T. Merle for useful discussions.
This work was supported by the Fonds de la Recherche Scientifique-FNRS under Grant No IISN 4.4502.19. 
L.S. and S.G. are senior FRS-F.N.R.S. research associates.

\bibliographystyle{aa}
\bibliography{astro.bib}


\begin{appendix}

\section{Comparison with observations in r/s-stars}

Figure~\ref{fig_fits} shows the individual best fits to the 14 r/s-stars of Table~\ref{table_fits} using our reference AGB model experiencing proton ingestion.

\begin{figure*}[h!]
 \begin{minipage}[c]{2\columnwidth}
\includegraphics[width=0.5\columnwidth]{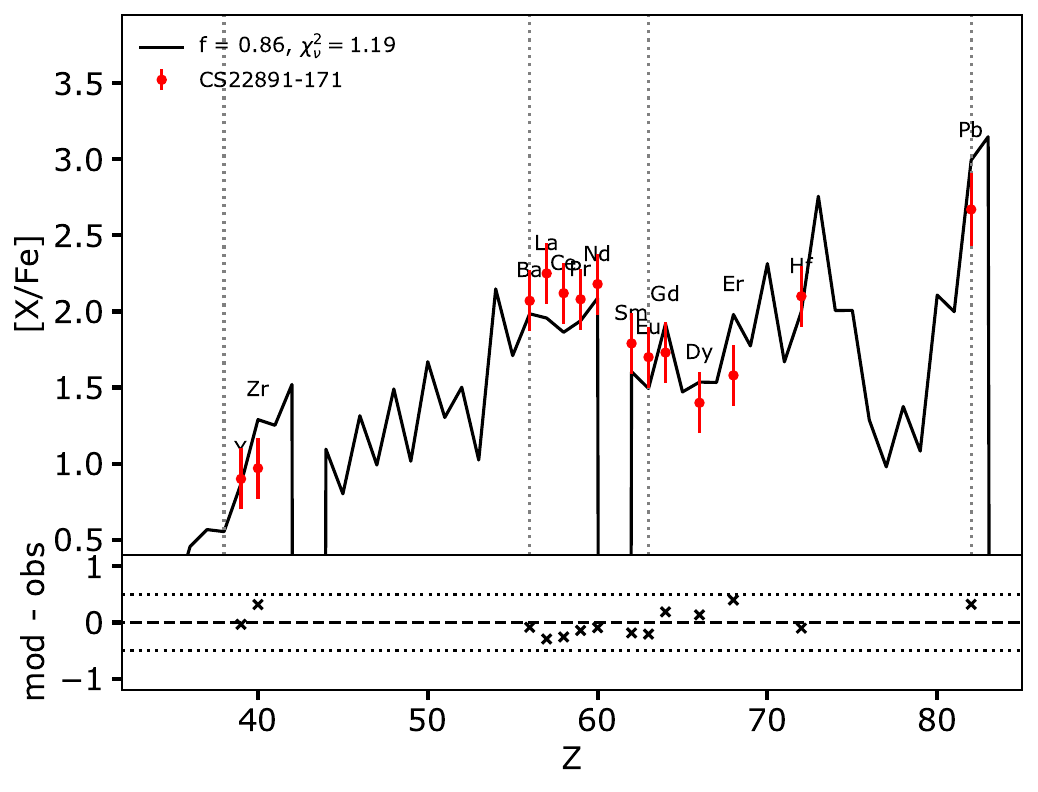}
\includegraphics[width=0.5\columnwidth]{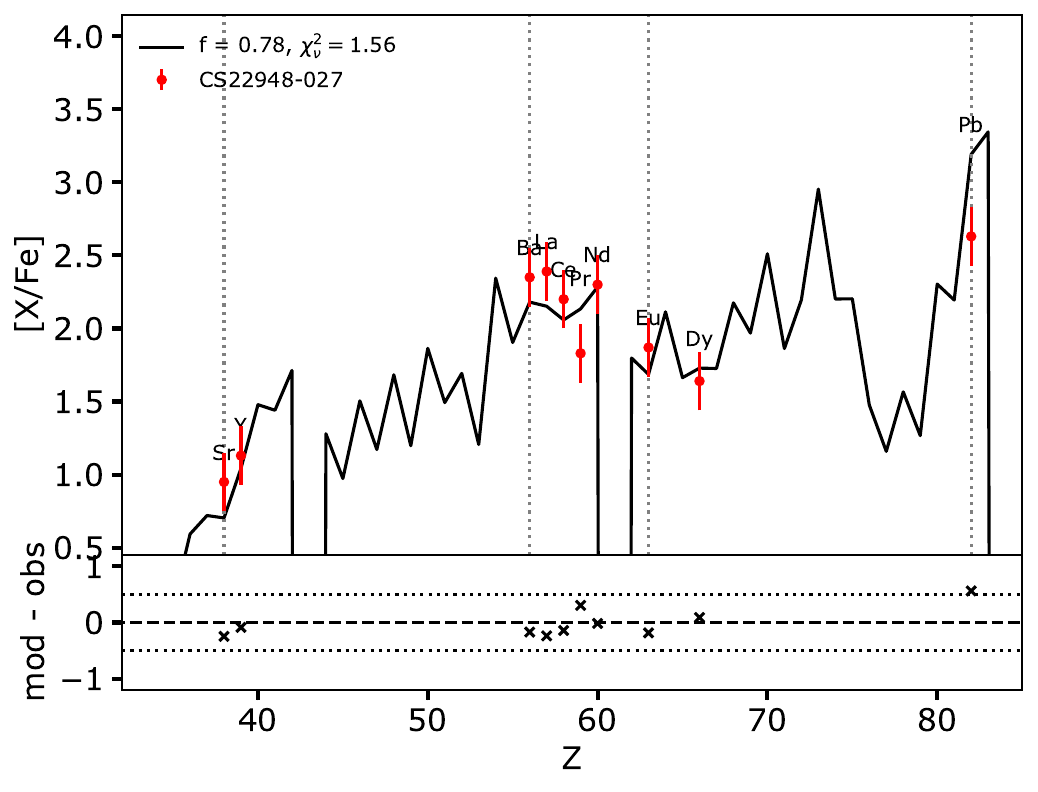}
  \end{minipage}
\begin{minipage}[c]{2\columnwidth}
\includegraphics[width=0.5\columnwidth]{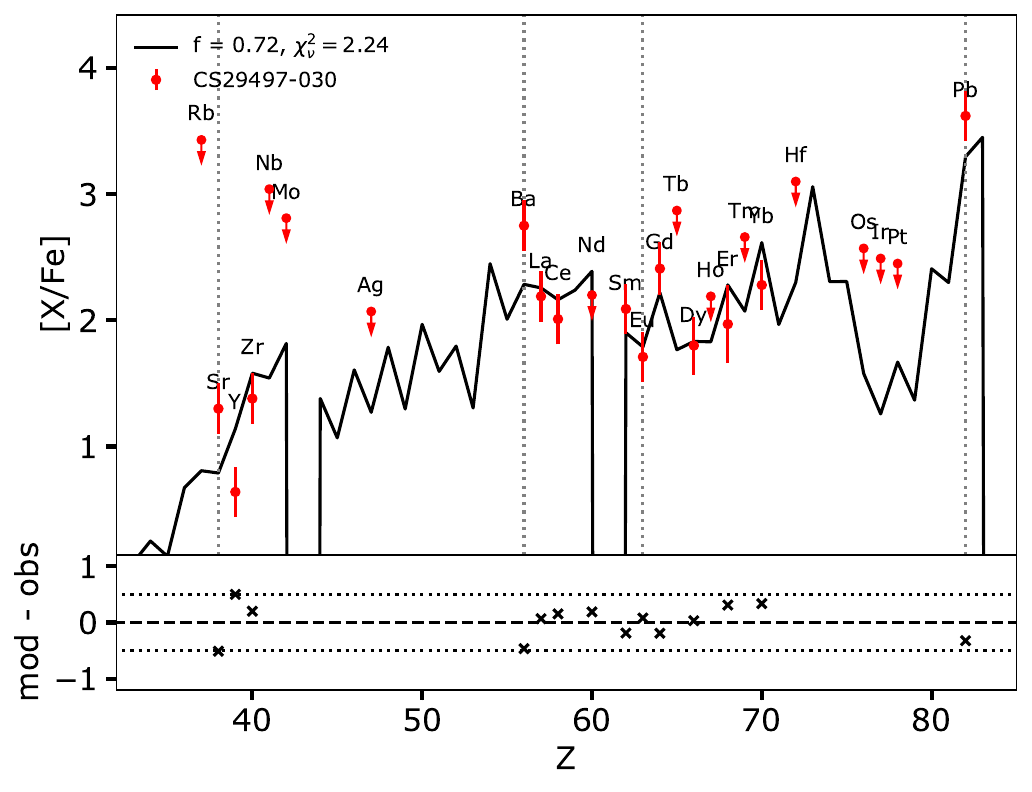}
\includegraphics[width=0.5\columnwidth]{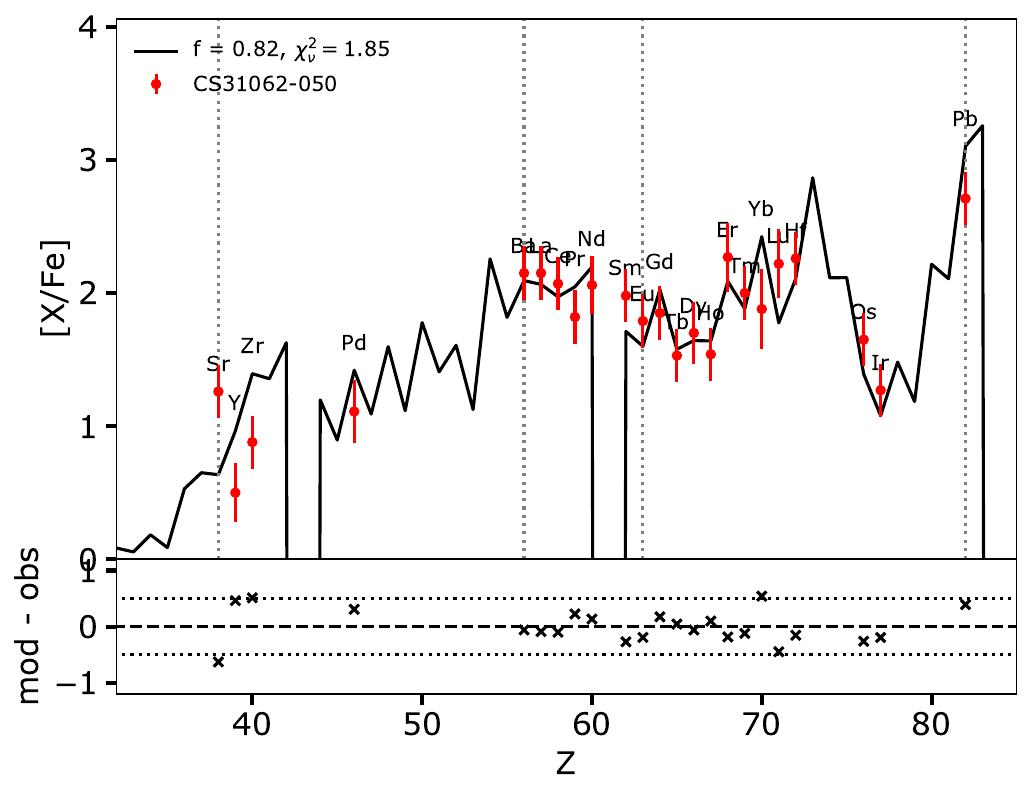}
  \end{minipage}
\begin{minipage}[c]{2\columnwidth}
\includegraphics[width=0.5\columnwidth]{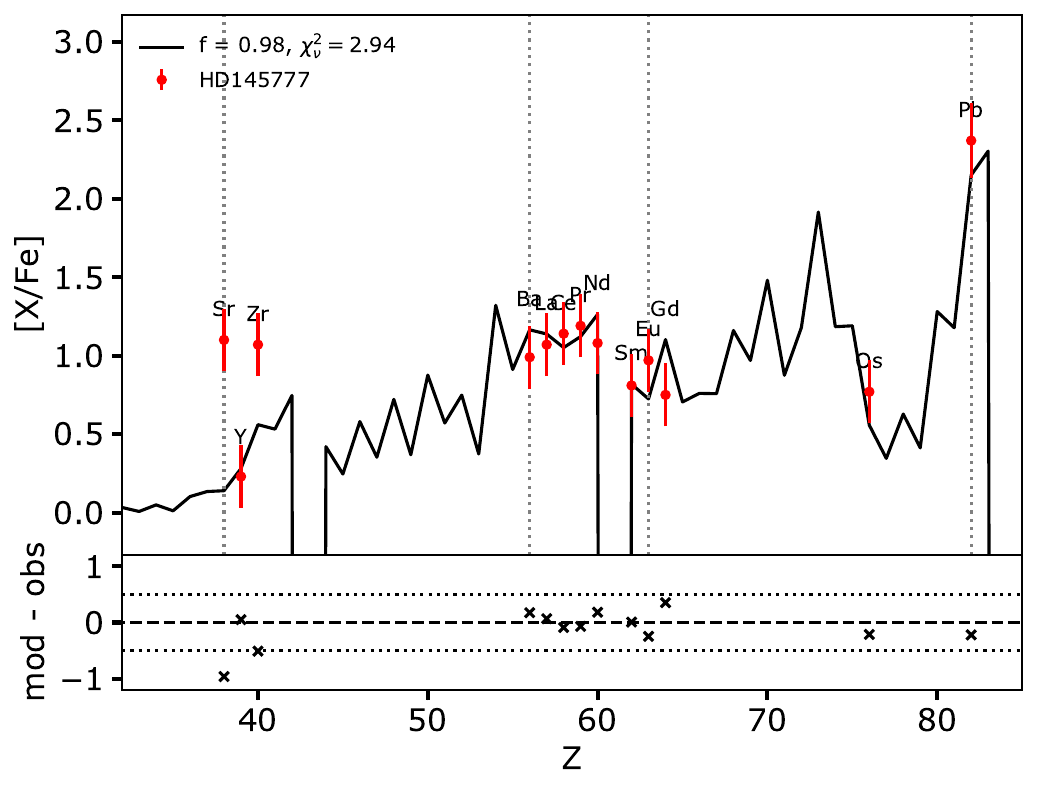}
\includegraphics[width=0.5\columnwidth]{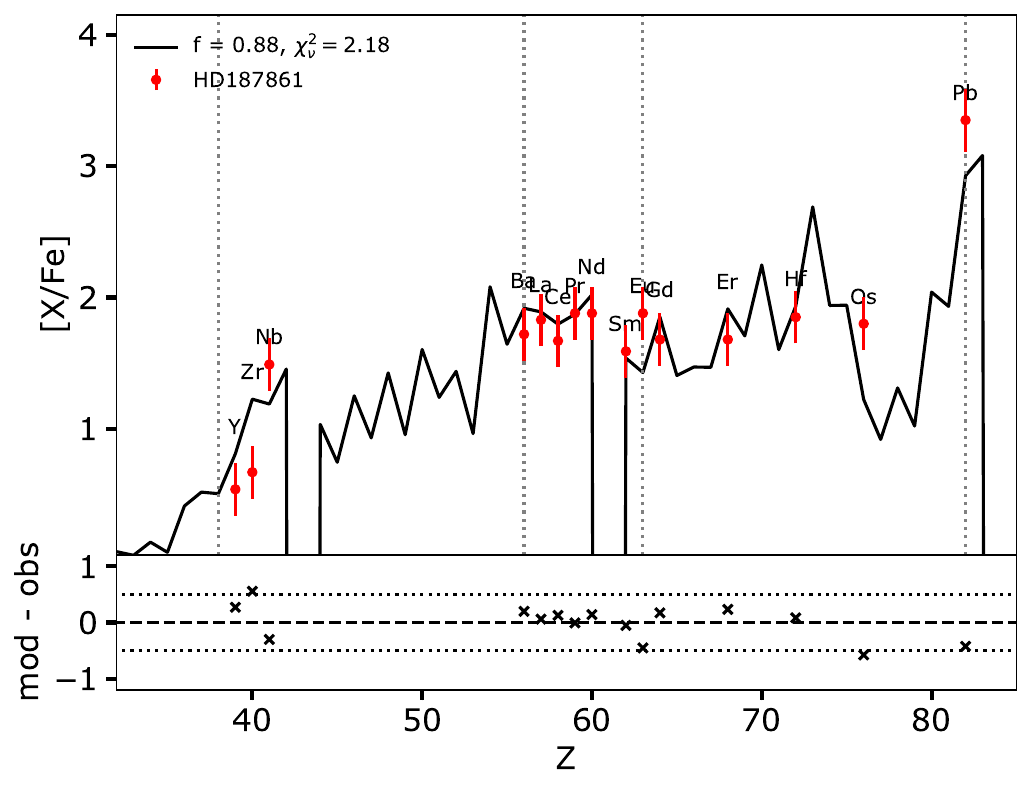}
  \end{minipage}
\caption{Best fits to the sample stars using the 1 \Msun\ AGB model with [Fe/H]~$=-2.5$ discussed throughout this paper (black patterns). The dilution factor $f$ (Eq.~\ref{eq_abdil}) and smallest $\chi_{\nu}^2$ value (Eq.~\ref{eq_chi}) are indicated. 
The abundance data are taken from the SAGA database \citep{suda08}.
}
\label{fig_fits}
\end{figure*}

\begin{figure*}[h]
   \ContinuedFloat
 \begin{minipage}[c]{2\columnwidth}
\includegraphics[width=0.5\columnwidth]{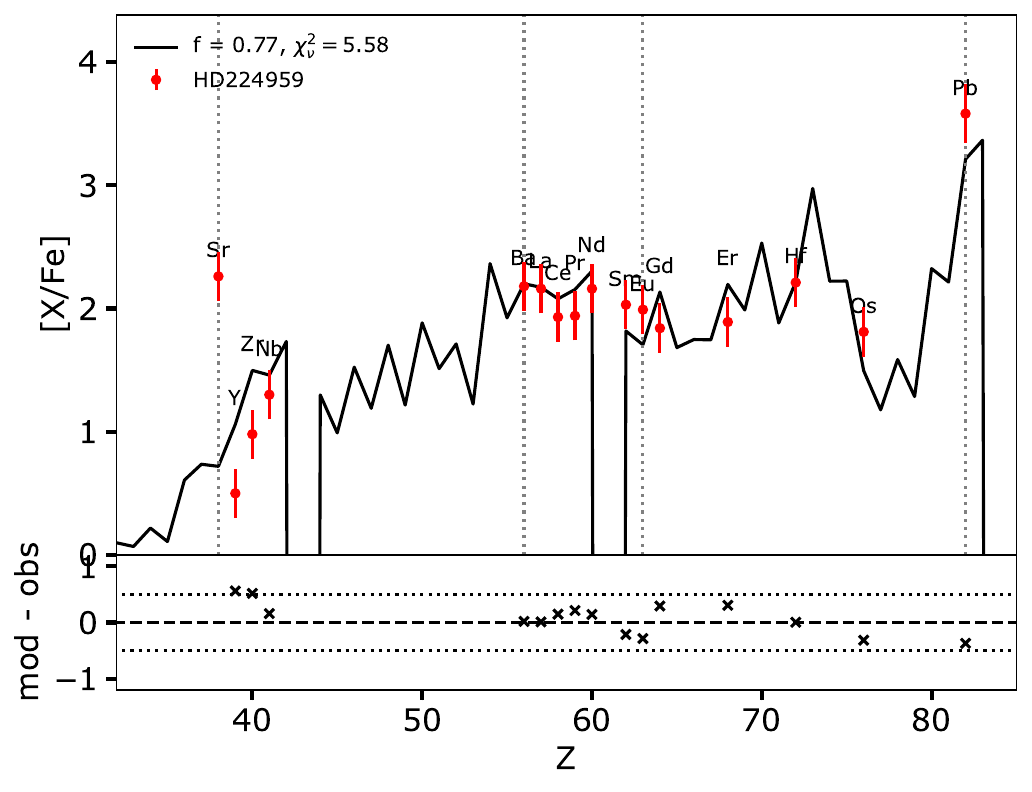}
\includegraphics[width=0.5\columnwidth]{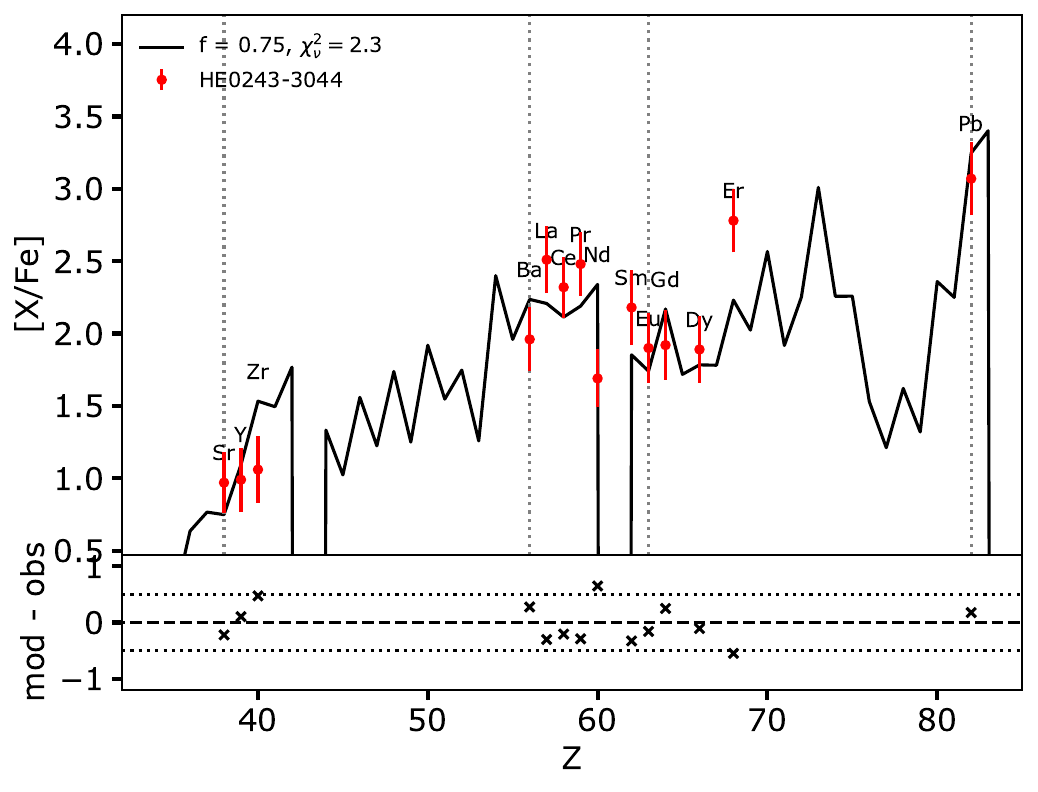}
  \end{minipage}
\begin{minipage}[c]{2\columnwidth}
\includegraphics[width=0.5\columnwidth]{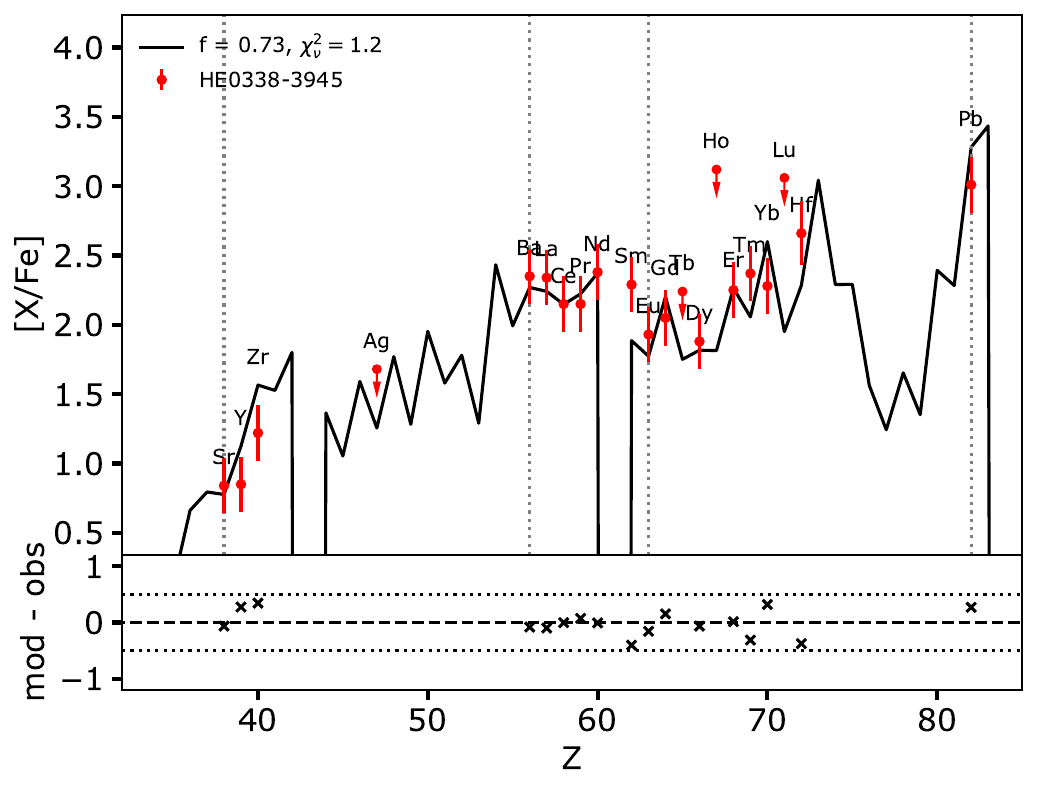}
\includegraphics[width=0.5\columnwidth]{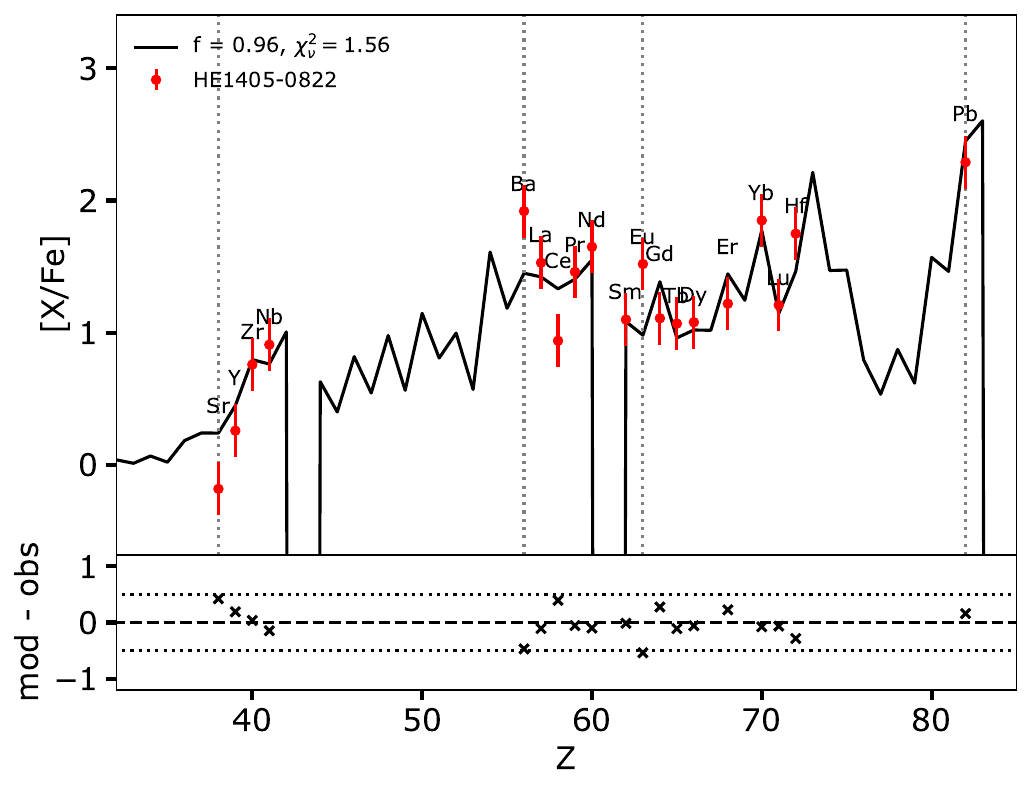}
  \end{minipage}
\begin{minipage}[c]{2\columnwidth}
\includegraphics[width=0.5\columnwidth]{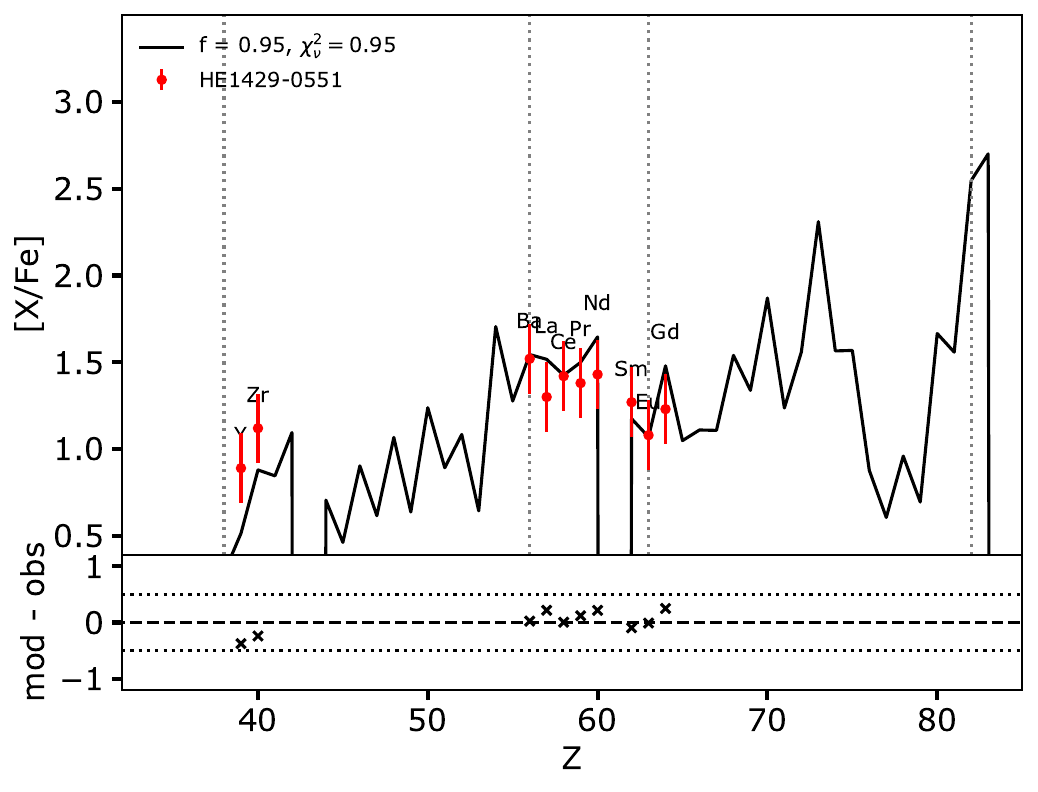}
\includegraphics[width=0.5\columnwidth]{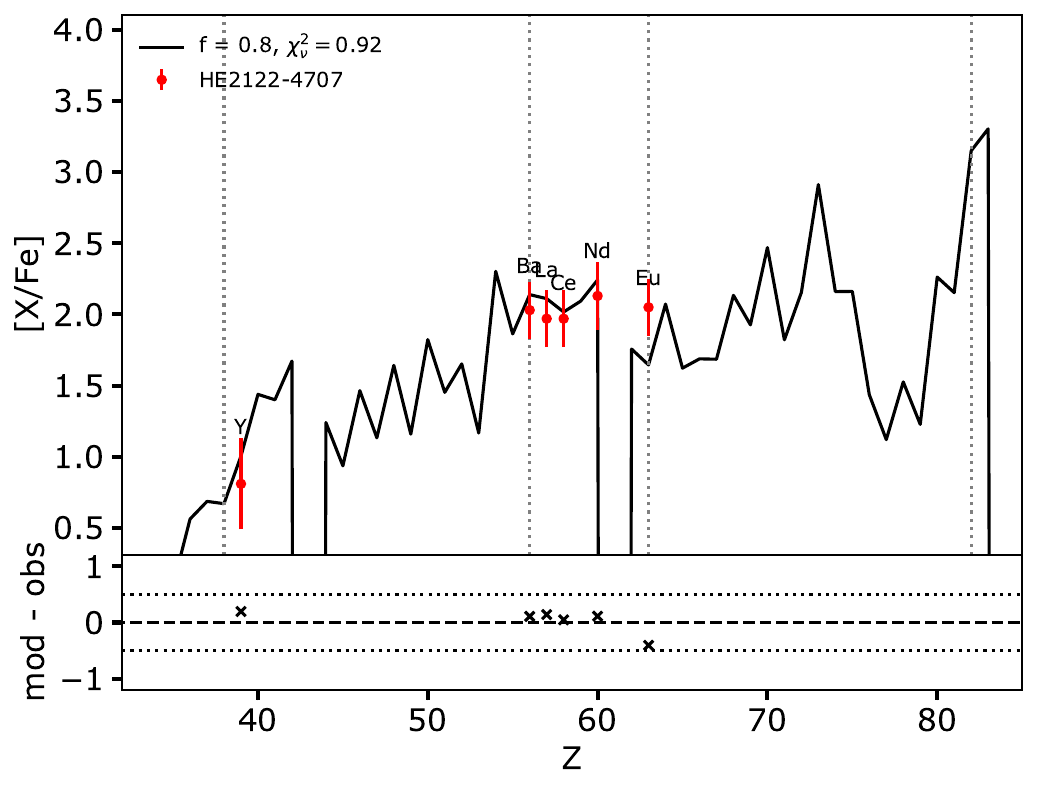}
  \end{minipage}
\caption{Continued.}
\end{figure*}

\begin{figure*}[h]
   \ContinuedFloat
 \begin{minipage}[c]{2\columnwidth}
\includegraphics[width=0.5\columnwidth]{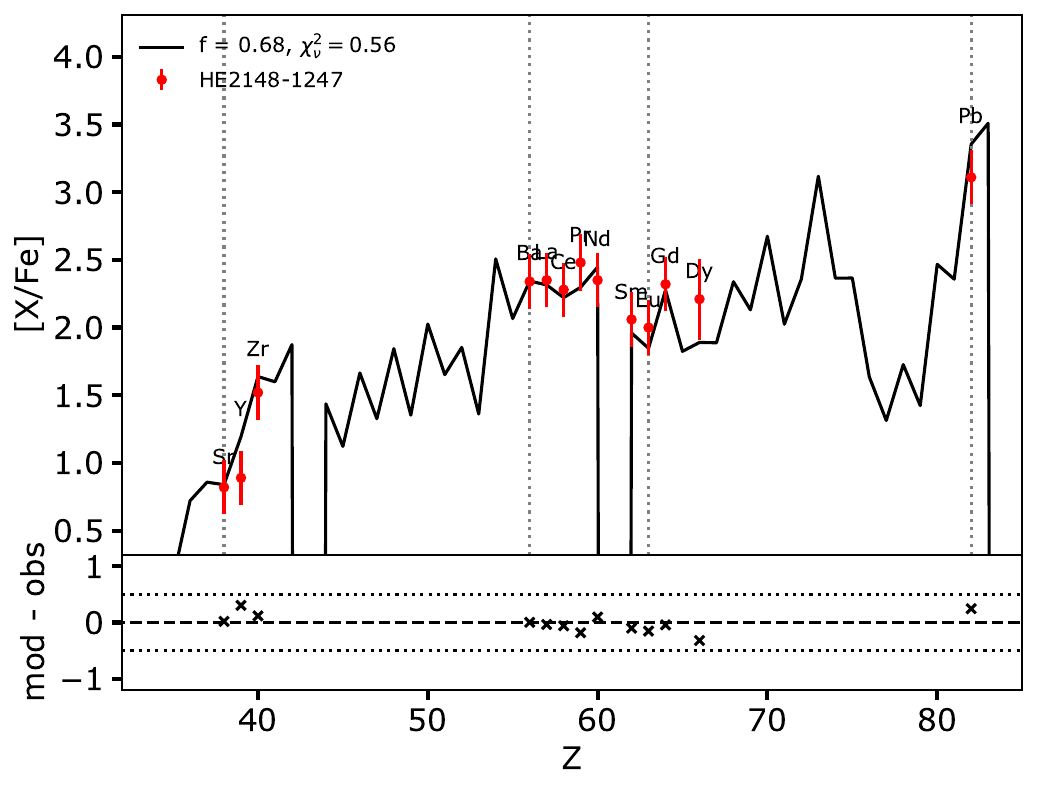}
\includegraphics[width=0.5\columnwidth]{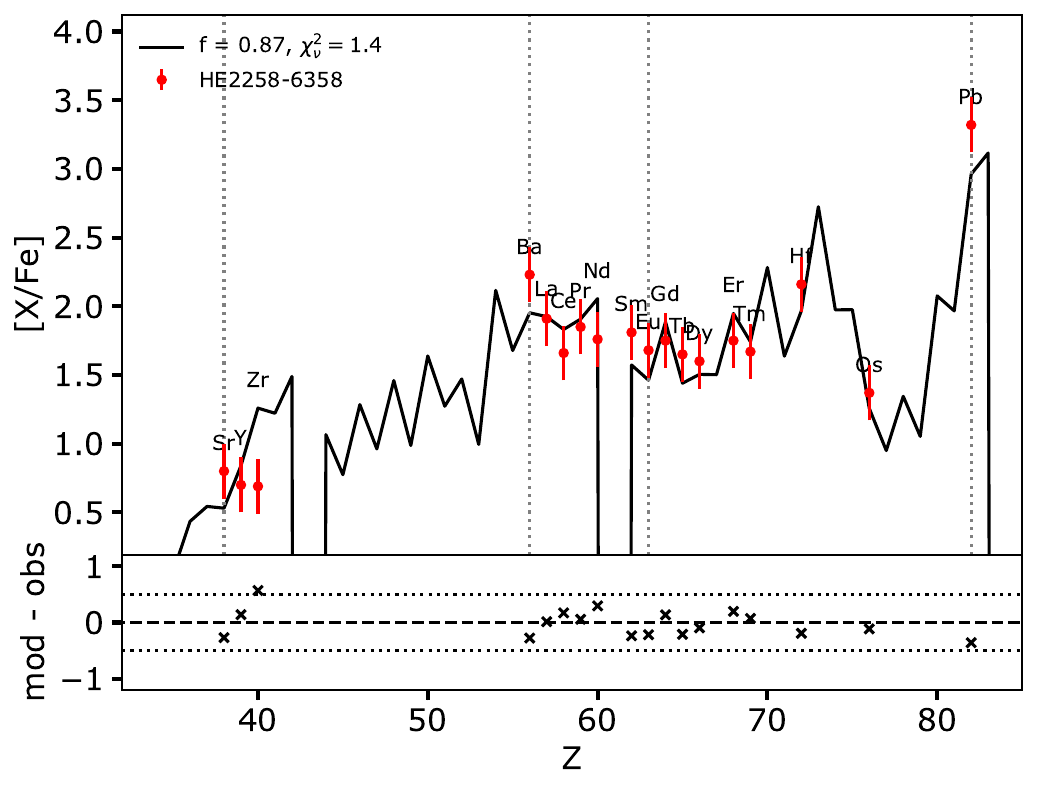}
  \end{minipage}
\caption{Continued.}
\end{figure*}

\end{appendix}

\clearpage
\newpage



\section*{Supplementary material}

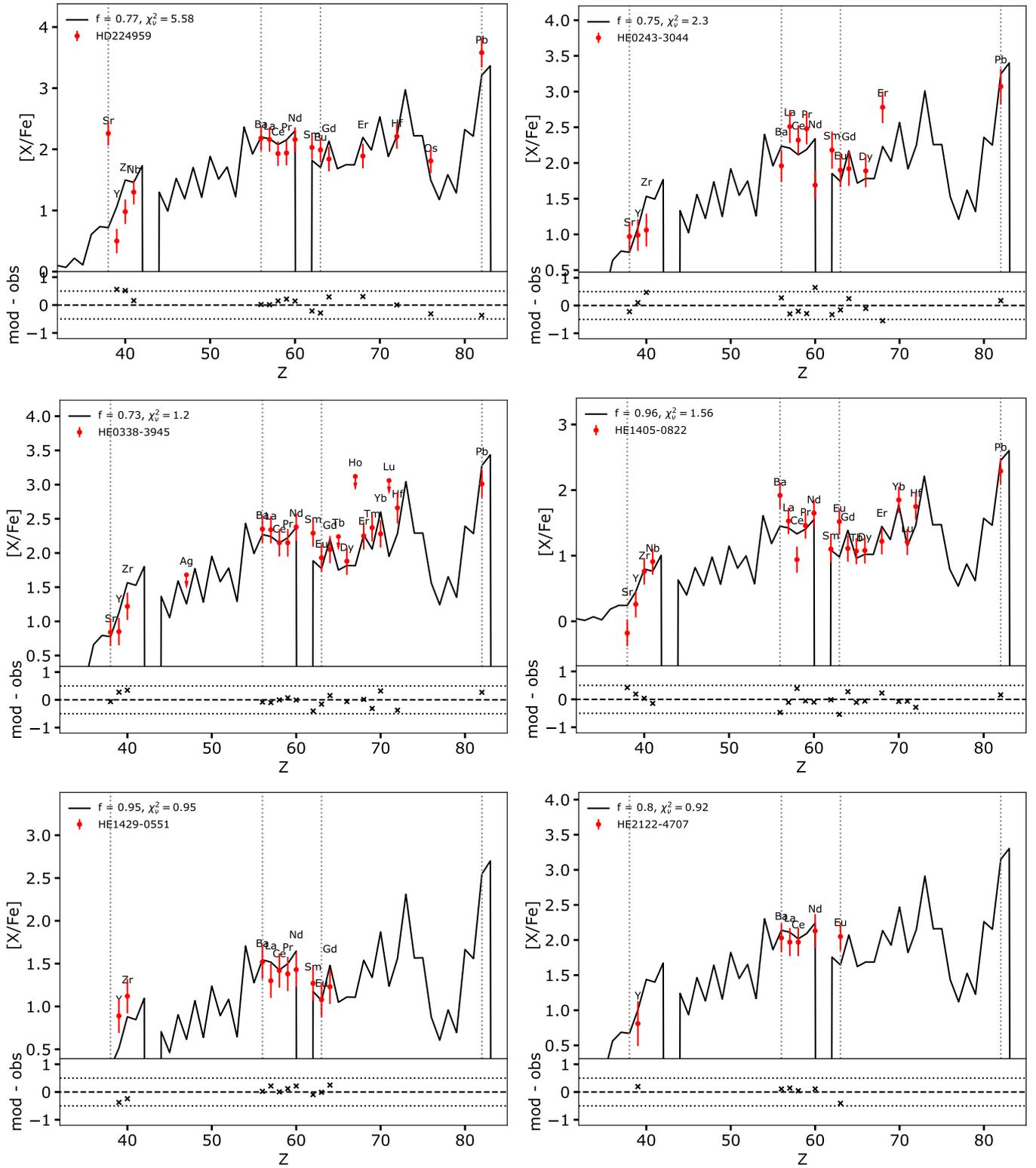

**Fig. A.1.** Continued.

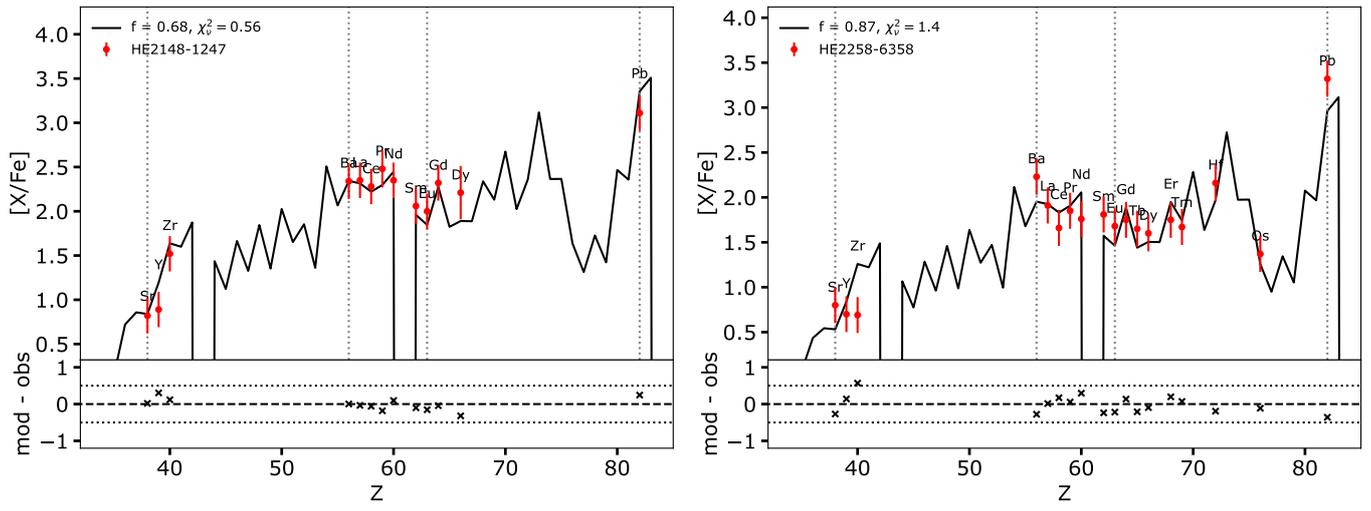

**Fig. A.1.** Continued.

# The intermediate neutron capture process

## I. Development of the i-process in low-metallicity low-mass AGB stars (Corrigendum)

A. Choplin, L. Siess, and S. Goriely

Institut d'Astronomie et d'Astrophysique, Université Libre de Bruxelles, CP 226, B-1050 Brussels, Belgium

June 7, 2022

**Key words.** nuclear reactions, nucleosynthesis, abundances – stars: AGB and post-AGB

We noticed that the dilution factors, $f$, reported in Table 2 of our paper (Choplin et al. 2021) were wrongly estimated. The new values, as provided in Table 1 of this erratum, are significantly larger than initially reported. We now find $0.68 < f < 0.98$, while in our original paper values of $0.15 < f < 0.26$ were quoted. The agreement between models and observation (in terms of chemical abundances) is not affected by the new values of the dilution factor, as seen in Table 1, where the updated $\chi^2_{\nu,\min}$ do not differ by more than 3 % from the previous values.

The only impact of this change is in the discussion of Sect. 6.5. In this section, we linked the dilution factor, $f$, to physical quantities related to the binary system properties (Eq. 10),

$$f = \frac{M_{\rm env}}{M_{\rm env} + M_{\rm acc}} = \frac{M_{\rm env}}{M_{\rm env} + \beta M_{\rm wind}},$$

where $M_{\rm env}$ is the envelope mass of the r/s star before the accretion episode, $M_{\rm acc}$ the mass accreted by the r/s star from the asymptotic giant branch (AGB) companion, $M_{\rm wind}$ the mass lost by the AGB star during its lifetime, and $\beta$ the accretion efficiency parameter. We highlighted that for $f \simeq 0.2$, the amount of mass that must be accreted ($M_{\rm acc}$) to account for the enrichment exceeds 1 $M_\odot$ for giant stars with deep convective envelopes ($M_{\rm env} \gtrsim 0.4\,M_\odot$). These giants represent 11 out of the 14 selected r/s stars, and the problem was that reaching such high values of $M_{\rm acc}$ cannot be physically achieved by wind accretion (Roche lobe overflow from an AGB star as a means to pollute the binary companion is unlikely as this process will most likely lead to a common envelope evolution).

With the updated values of $f$ (higher by a factor of $4-5$), the 11 giant stars now have $0.75 < f < 0.98$, which leads to a much more reasonable accreted mass of $8 \times 10^{-3} < M_{\rm acc}/M_\odot < 0.125$ with a mean of 0.06 $M_\odot$. The accretion efficiency parameter, $\beta$, now ranges between 0.015 and 0.25 for giant stars. This is compatible with 3D hydrodynamical simulations of wind mass transfer in AGB binary systems (e.g. Saladino et al. 2019).

Overall, the correction of the dilution factors yields more realistic accreted mass and accretion efficiency parameters. It gives more credit to the AGB wind accretion scenario for explaining the high level of enrichment noticed in most r/s stars. Therefore, this gives even more confidence in low-metallicity AGB stars as a viable site for i-process nucleosynthesis.

**Table 1.** Characteristics and fit parameters of the 14 selected r/s stars. The quantity $N$ is the number of data points between Zn ($Z = 30$) and Bi ($Z = 83$), $d_{\rm RMS}$ the average distance to the solar scaled r-process (Eq. 5), $\chi^2_{\nu,\min}$ the minimum reduced $\chi^2_\nu$ value (Eq. 7), and $f$ the dilution factor (Eq. 8).

| Star | [Fe/H] | $N$ | $d_{\rm RMS}$ | $\chi^2_{\nu,\min}$ | $f$ | Ref |
|---|---|---|---|---|---|---|
| CS22891-171 | -2.5 | 14 | 0.79 | 1.19 | 0.86 | 1, 2 |
| CS22948-027 | -2.26 | 10 | 0.60 | 1.56 | 0.78 | 3, 4, 5 |
| CS29497-030 | -2.52 | 13 | 0.77 | 2.24 | 0.72 | 6, 7, 8 |
| CS31062-050 | -2.48 | 23 | 0.76 | 1.85 | 0.82 | 9, 10, 11, 12, 13 |
| HD145777 | -2.32 | 13 | 0.79 | 2.94 | 0.98 | 1 |
| HD187861 | -2.6 | 15 | 0.70 | 2.18 | 0.88 | 1, 2 |
| HD224959 | -2.36 | 16 | 0.80 | 5.58 | 0.77 | 1, 2 |
| HE0243-3044 | -2.58 | 14 | 0.79 | 2.3 | 0.75 | 14 |
| HE0338-3945 | -2.43 | 17 | 0.76 | 1.2 | 0.73 | 15 |
| HE1405-0822 | -2.37 | 19 | 0.62 | 1.56 | 0.96 | 16 |
| HE1429-0551 | -2.7 | 10 | 0.70 | 0.95 | 0.95 | 1 |
| HE2122-4707 | -2.47 | 6 | 0.57 | 0.92 | 0.8 | 17 |
| HE2148-1247 | -2.35 | 13 | 0.80 | 0.56 | 0.68 | 17, 18 |
| HE2258-6358 | -2.67 | 18 | 0.75 | 1.40 | 0.87 | 19 |

**References.** 1 - Karinkuzhi et al. (2021); 2 - Masseron et al. (2010); 3 - Aoki et al. (2002); 4 - Aoki et al. (2007); 5 - Barbuy et al. (2005); 6 - Roederer et al. (2014); 7 - Ivans et al. (2005); 8 - Sivarani et al. (2004); 9 - Lai et al. (2007); 10 - Aoki et al. (2002); 11 - Lai et al. (2004); 12 - Johnson & Bolte (2004); 13 - Aoki et al. (2006); 14 - Hansen et al. (2015); 15 - Jonsell et al. (2006); 16 - Cui et al. (2013); 17 - Cohen et al. (2013); 18 - Cohen et al. (2003); 19 - Placco et al. (2013).



\end{document}